\documentclass[fleqn,usenatbib]{mnras}
\usepackage[T1]{fontenc}
\usepackage{newtxtext,newtxmath}
\usepackage{graphicx,amsmath,booktabs,xcolor,siunitx,hyperref}
\usepackage{multirow}
\usepackage{comment}
\newcommand{\codeName}{\textsc{ANRe-M1}}

\title[GPU-accelerated supernova code]{ANRe-M1: a GPU-accelerated numerical relativity code with multi-energy M1 neutrino transport}
\author[T. Kuroda \& M. Shibata]{Takami Kuroda,$^{1}$\thanks{E-mail: takami.kuroda@aei.mpg.de}
and Masaru Shibata$^{1,2}$\\
$^{1}$Max-Planck-Institut f{\"u}r Gravitationsphysik, Am M{\"u}hlenberg 1, D-14476 Potsdam-Golm, Germany\\
$^{2}$Center for Gravitational Physics and Quantum-Information,
Yukawa Institute for Theoretical Physics, Kyoto University, Kyoto, 606-8502, Japan}
\date{Accepted XXX. Received YYY; in original form ZZZ}
\pubyear{2026}

\begin{document}
\label{firstpage}
\pagerange{\pageref{firstpage}--\pageref{lastpage}}
\maketitle

\begin{abstract}
We present ANRe-M1 (\textbf{A}ccelerated \textbf{N}umerical \textbf{Re}lativity code with multi-energy \textbf{M1} neutrino transport), a performance-portable, GPU-accelerated code for multidimensional core-collapse supernova simulations in full general relativity.
ANRe-M1 builds on our original CPU-based Fortran framework and is implemented in C++ using Kokkos.
It retains the underlying general-relativistic neutrino-radiation-hydrodynamics algorithms while redesigning the data layout, parallel decomposition, and memory-access patterns for accelerator-resident execution.
The spacetime sector employs the BSSN--Z4c formulation and is coupled to relativistic hydrodynamics and energy-dependent M1 neutrino transport.
We validate the implementation through a suite of hydrodynamic, radiation-transport, and dynamical-spacetime tests, together with comparisons to results from an earlier code-comparison study.
We further demonstrate its applicability with a representative three-dimensional stellar-collapse computation including multi-energy neutrino transport.
For the benchmark and computing systems considered here, ANRe-M1 runs approximately an order of magnitude faster on a cluster of AMD MI300A accelerated processing units than the Fortran version on a CPU cluster.
When the allocation is increased from 2 to 128 MI300A APUs, the aggregate throughput rises by a factor of 39.7, corresponding to a weak-scaling efficiency of 62 per cent relative to the two-APU case.
These results establish ANRe-M1 as an efficient and portable framework for long-term, multidimensional supernova simulations on current and forthcoming accelerator-based supercomputers.
\end{abstract}


\section{Introduction}
\label{sec:introduction}

Core-collapse supernovae mark the violent deaths of stars born with masses above approximately \(8\,{\rm M}_{\odot}\), leaving neutron stars or black holes and injecting kinetic energy and newly synthesized elements into their surroundings.
During collapse, the central core reaches nuclear density, stiffens, and launches a shock at core bounce, but the shock rapidly loses energy through nuclear dissociation and neutrino emission and typically stalls inside the iron core.
In the delayed neutrino-heating mechanism, electron-flavour neutrinos emitted by the nascent proto-neutron star deposit a fraction of their energy behind the stalled shock and can revive it to produce an explosion \citep[e.g.][]{Janka2025_review,Burrows2021_review,Mezzacappa2023,Fryer2023,Yamada2024}.
Whether this revival occurs is controlled by the nonlinear interaction of gravity, hydrodynamics, neutrino transport, and nuclear microphysics across a wide range of spatial and temporal scales.

Multidimensional dynamics are an essential part of this mechanism rather than a small correction to a spherical background.
Neutrino-driven convection increases the dwell time of accreted matter in the gain region and thereby enhances neutrino heating \citep{Radice2016,Burrows2020}.
The standing accretion shock instability drives large-scale sloshing and spiral motions of the stalled shock \citep{Summa2016,Walk2018}, while turbulent Reynolds stresses provide additional support against the upstream ram pressure \citep{Couch2015,Nagakura2019,Radice2018}.
Rotation can modify the post-shock flow and the critical condition for runaway \citep{Summa2018,Takiwaki2016}.
Finally, non-spherical perturbations inherited from convective silicon- and oxygen-shell burning can seed post-shock turbulence and facilitate shock revival \citep{Mueller2017,NagakuraAsym2019,Vartanyan2022}.

General relativity is also important because the deeper relativistic gravitational potential produces a more compact and hotter proto-neutron star, modifies the accretion flow, and changes the luminosities and spectra of the emitted neutrinos \citep{Mueller2012,OConnorCouch2018}.
These differences feed back on the gain-region heating conditions and can materially affect the proximity to shock revival; direct comparisons with Newtonian gravity have found more favourable conditions for neutrino-driven explosions when relativistic gravity is included \citep{OConnorCouch2018}.
Multidimensional full-GR computations coupled to multi-energy neutrino moment transport now follow collapse, bounce, and the stalled-shock phase self-consistently \citep{Kuroda2016,Roberts2016}.
For very massive and compact progenitors, a relativistic treatment is also indispensable for capturing the second collapse of the proto-neutron star and black-hole formation; a three-dimensional full-GR simulation by \citet{Kuroda2018} found a short-lived neutrino-driven shock revival shortly before black-hole formation, while \citet{Chan2018} followed an aborted neutrino-driven explosion through black-hole formation and subsequent fallback.
Quantitative predictions therefore require simulations, ideally of sufficiently long duration, that combine multidimensional dynamics, energy-dependent neutrino transport, and an adequate treatment of relativistic gravity.

Neutrino radiation is a fundamental ingredient of this supernova picture.
Through deleptonization during collapse and throughout the post-bounce evolution, it governs the thermodynamic and compositional properties of the proto-neutron star (PNS); through energy deposition behind the stalled shock, it also underpins the standard neutrino-heating mechanism for shock revival.
Three-dimensional neutrino-radiation-hydrodynamics computations incorporating these ingredients are consequently becoming the standard for state-of-the-art studies of the explosion mechanism.
The codes used in this field span a hierarchy of approximations in both gravity and neutrino transport.
\textsc{Prometheus-Vertex} employs Newtonian hydrodynamics with an effective relativistic gravitational potential and energy-dependent two-moment transport closed by a model Boltzmann equation, with multidimensional transport treated in the ray-by-ray-plus approximation \citep{Bollig2021}.
\textsc{CoCoNuT-Vertex} combines the same transport approach with general-relativistic hydrodynamics and an extended conformal-flatness approximation for the spacetime \citep{Mueller2012}.
\textsc{Fornax} solves multidimensional, multi-group comoving-frame moment equations with an M1 closure and is commonly used for three-dimensional supernova computations with an approximate treatment of relativistic gravity \citep{Skinner2019,Burrows2020}.
\textsc{Alcar} also evolves energy-dependent zeroth and first angular moments with an algebraic closure, including the leading velocity-dependent terms, and has been applied with Newtonian or effective-relativistic gravity \citep{Just2015}.
\textsc{Chimera} combines multidimensional hydrodynamics with detailed multi-group neutrino transport in a ray-by-ray-plus formulation and includes relativistic corrections through an effective gravitational potential \citep{Bruenn2020}.
At the most complete transport level, multidimensional discrete-ordinate Boltzmann solvers have been developed in special and full general relativity, but their high dimensionality makes routine three-dimensional production computations exceptionally demanding \citep{Nagakura2018,Akaho2021}.
Our earlier code instead evolves the dynamical spacetime in full \(3+1\) general relativity and treats multi-energy neutrino radiation with an M1 moment scheme \citep{Kuroda2016}.

Increasing computational resources and improvements in numerical methods have also extended three-dimensional simulations well beyond shock revival.
For example, \citet{Mueller2017} followed a \({\sim}18\,{\rm M}_{\odot}\) model with multi-group neutrino hydrodynamics for more than \(2\) s after bounce.
In \citet{Bollig2021}, the final \(7\) min of convective oxygen-shell burning were evolved in three dimensions before collapse, and the ensuing supernova was followed to more than \(7\) s after bounce.
\citet{Nakamura2019Long} evolved two three-dimensional \(11.2\,{\rm M}_{\odot}\) models to approximately \(1.1\) s after bounce to investigate explosion asymmetries and neutron-star kicks.
Other long-duration studies include the low-mass progenitor models of \citet{Stockinger2020Long}, which were followed from collapse through shock breakout, and the \textsc{Fornax} programme, which covered a wide progenitor mass range in three-dimensional simulations \citep{Burrows2020} and subsequently evolved twelve models spanning \(9\)--\(60\,{\rm M}_{\odot}\) for multi-second studies of neutrino-driven winds \citep{Wang2023}.
Such long-duration computations are necessary because explosion energies, remnant masses and kicks, anisotropic accretion, neutrino-driven outflows, and nucleosynthetic conditions can continue to evolve for seconds after the onset of explosion.

These advances come at substantial computational cost.
A three-dimensional computation must evolve the fluid and spacetime together with multi-energy radiation moments for several neutrino species, while repeatedly evaluating stiff neutrino--matter interaction terms.
Extending such simulations to higher resolution, broader progenitor surveys, and post-bounce times of several seconds remains challenging on conventional CPU-only systems.
At the same time, contemporary leadership-class supercomputers increasingly obtain most of their floating-point throughput and memory bandwidth from GPUs and other accelerators.
Moving supernova codes to these architectures can therefore shorten the total simulation time and, when the implementation uses the accelerator efficiently, improve energy efficiency.
Doing so requires abundant fine-grained parallelism, suitable data layouts, and careful control of data movement rather than a simple recompilation of a CPU-oriented code.

Several recent astrophysical codes have already moved in this direction.
\textsc{Phoebus} is a GPU-first, performance-portable general-relativistic radiation-magnetohydrodynamics code built on \textsc{Parthenon} and Kokkos, with both Monte Carlo and moment methods for neutrino transport and demonstrated scaling to more than \(500\) NVIDIA H100 GPUs \citep{Barker2024}.
Its moment solver is gray, evolving energy-integrated zeroth and first moments for three species, whereas its Monte Carlo solver retains spectral information; the published verification consists of transport tests rather than an end-to-end collapse-to-post-bounce comparison \citep{Barker2024}.
For dynamical-spacetime numerical relativity, \textsc{GRaM-X} combines Z4c spacetime evolution and Valencia general-relativistic magnetohydrodynamics with the GPU-capable CarpetX/\textsc{AMReX} adaptive-mesh infrastructure \citep{Shankar2023}.
Recent three-dimensional magnetorotational core-collapse simulations with \textsc{GRaM-X} evolved the dynamical spacetime on GPUs, but employed the computationally inexpensive M0 neutrino approximation rather than multi-energy M1 transport \citep{Shankar2026}.
The numerical-relativity module of \textsc{AthenaK} likewise uses Kokkos for performance portability, evolves the Z4c system, and has demonstrated large-scale execution on GPU-based exascale systems \citep{Zhu2025}.
\textsc{AsterX} is an open-source GPU-accelerated general-relativistic magnetohydrodynamics code for dynamical spacetimes that is built within the Einstein Toolkit and uses CarpetX/\textsc{AMReX} for adaptive mesh refinement \citep{Kalinani2025}.
More recently, \textsc{SACRA-K} ported the established Fortran \textsc{SACRA-MPI} compact-binary code to C++ and Kokkos while retaining BSSN evolution with Z4c constraint propagation, general-relativistic hydrodynamics, and mesh refinement \citep{Han2026}.
\textsc{GRACE} provides another open-source Kokkos-based framework, coupling Z4c spacetime evolution to ideal general-relativistic magnetohydrodynamics on fixed or adaptively refined meshes \citep{Musolino2026}.
\textsc{MHDuet} couples the full Einstein equations and general-relativistic magnetohydrodynamics to a gray, energy-integrated three-species M1 scheme on AMReX GPUs, but its published validation focuses on neutron-star configurations rather than stellar core collapse \citep{Palenzuela2025}.
Conversely, \textsc{thornado+Flash-X} provides spectral, six-species two-moment transport with GPU offloading and has demonstrated spherically symmetric and axisymmetric core-collapse simulations, including agreement with \textsc{Chimera}; it employs self-gravitating hydrodynamics with observer corrections accurate to \(\mathcal{O}(v/c)\), rather than a dynamical full-GR spacetime \citep{Endeve2026}.
These developments demonstrate the maturity of GPU-accelerated relativistic hydrodynamics and dynamical-spacetime evolution, but most of the numerical-relativity frameworks in this list do not include the multi-species, multi-energy neutrino transport required for self-consistent core-collapse supernova simulations.
To our knowledge, the published literature therefore does not yet contain a GPU-accelerated code that combines a dynamical full-GR spacetime with multi-species, multi-energy M1 transport.

In \citet{Kuroda2016}, we introduced a multidimensional, multi-energy neutrino-radiation-hydrodynamics code in full general relativity and applied it to the collapse of massive stars.
The same framework was subsequently used to demonstrate a three-dimensional magnetorotational explosion whose broader outflow was jointly supported by magnetic pressure and neutrino heating \citep{KurodaArcones2020} and to connect rotation-driven non-axisymmetric dynamics with correlated gravitational-wave and neutrino variability \citep{Shibagaki2021}.
It was further applied to investigate a first-order QCD phase transition and the formation of neutron stars, hybrid stars, and black holes \citep{Kuroda2022QCD}.
More recent applications include radiation-magnetohydrodynamics simulations of black-hole and relativistic-jet formation in rotating massive stars \citep{KurodaShibata2024} and three-dimensional simulations of rotating magnetized white-dwarf collapse and its multimessenger signals \citep{Kuroda2025AIC}.
That framework was written in Fortran and designed primarily for conventional CPU-based systems.
In the present work, we report \codeName, a performance-portable update of that code written in C++ using Kokkos for GPU-accelerated core-collapse supernova simulations.
The new implementation retains the general-relativistic radiation-hydrodynamics formulation of the original code, while replacing the spacetime sector with the BSSN--Z4c formulation and restructuring the principal computational kernels, data layouts, and memory access patterns for heterogeneous execution.
It evolves a dynamical spacetime together with general-relativistic hydrodynamics and multi-species, multi-energy M1 neutrino transport using an arbitrary number \(N_{\varepsilon}\) of neutrino-energy groups.
Our goals are to verify agreement with the established Fortran implementation, quantify performance and scaling on an AMD MI300A APU cluster, and provide a maintainable foundation for future high-resolution and long-duration supernova simulations.
For the workload studied here, the new implementation runs approximately an order of magnitude faster on the MI300A APU cluster than the Fortran implementation on the CPU cluster.
The benchmark definition and the limitations of this cross-platform comparison are given in Section~\ref{sec:performance}.

The remainder of this paper is organized as follows.
Section~\ref{sec:formulation} summarizes the spacetime and neutrino-radiation-hydrodynamics formulation.
Section~\ref{sec:implementation} describes the Kokkos implementation.
Section~\ref{sec:verification} presents code verification.
Section~\ref{sec:application} gives a representative core-collapse application, Section~\ref{sec:performance} presents performance measurements, and Section~\ref{sec:conclusions} summarizes our conclusions.

\section{Formalism}
\label{sec:formulation}

The radiation-hydrodynamics formulation is essentially the same as that of \citet{Kuroda2016}.
The main change in the continuum equations is the replacement of the Baumgarte--Shapiro--Shibata--Nakamura (BSSN) formulation \citep{Shibata1995,Baumgarte1999} used in the original Fortran-based code by the conformal and constraint-propagation Z4 (Z4c) formulation.
We summarize both sectors below, using geometrized units, $G=c=1$, where $G$ and $c$ are the gravitational constant and speed of light, respectively.  Greek indices run over spacetime components and Latin indices over spatial components.

\subsection{BSSN--Z4c}
\label{sec:z4c}

We write the spacetime metric in the standard $3+1$ form,
\begin{equation}
 {\rm d}s^2=-\alpha^2{\rm d}t^2
 +\gamma_{ij}({\rm d}x^i+\beta^i{\rm d}t)
              ({\rm d}x^j+\beta^j{\rm d}t),
\label{eq:line_element}
\end{equation}
where $\alpha$, $\beta^i$, and $\gamma_{ij}$ are the lapse, shift, and
spatial metric, respectively.  The future-directed unit normal to a
constant-time hypersurface is
$n^\mu=(\alpha^{-1},-\beta^i/\alpha)$.  The matter sources of the
Einstein equations are defined by projections of the total stress-energy tensor, $T_{\rm total}^{\mu\nu}$, as
\begin{equation}
 {\cal E}=n_\mu n_\nu T_{\rm total}^{\mu\nu},\qquad
 {\cal S}_i=-\gamma_{i\mu}n_\nu T_{\rm total}^{\mu\nu},\qquad
 {\cal S}_{ij}=\gamma_{i\mu}\gamma_{j\nu}T_{\rm total}^{\mu\nu},
\label{eq:adm_sources}
\end{equation}
and ${\cal S}\equiv\gamma^{ij}{\cal S}_{ij}$.  Both the fluid and
energy-integrated neutrino moments contribute to these projections.

Following the Z4c formulation \citep{Bernuzzi2010,Hilditch2013}, we
introduce the conformal variables used in the code,
\begin{equation}
 W\equiv\gamma^{-1/6},\qquad
 \widetilde{\gamma}_{ij}\equiv W^2\gamma_{ij},\qquad
 \widetilde{A}_{ij}\equiv
 W^2\left(K_{ij}-\frac{1}{3}\gamma_{ij}K\right),
\label{eq:z4c_variables}
\end{equation}
where $\gamma=\det(\gamma_{ij})$ and $K=\gamma^{ij}K_{ij}$.  The
remaining Z4c variables are the scalar constraint $\Theta$, the
constraint-modified trace
\begin{equation}
 \widehat{K}\equiv K-2\Theta,
\end{equation}
and the independently evolved conformal connection functions
\begin{equation}
 \widetilde{\Gamma}^{i}
 \equiv\widetilde{\gamma}^{jk}\widetilde{\Gamma}^{i}{}_{jk}
 =-\partial_j\widetilde{\gamma}^{ij}.
\label{eq:z4c_gamma}
\end{equation}
The physical spatial metric is therefore recovered as $\gamma_{ij}=W^{-2}\widetilde{\gamma}_{ij}$.

Defining $\partial_\perp\equiv\partial_t-\beta^k\partial_k$, the
evolution equations implemented in the code are
\begin{align}
 \partial_\perp W={}&\frac{1}{3}W
 \left[\alpha(\widehat K+2\Theta)-\partial_k\beta^k\right],
\label{eq:z4c_w}\\
 \partial_\perp\widetilde{\gamma}_{ij}={}&
 -2\alpha\widetilde A_{ij}
 +2\widetilde{\gamma}_{k(i}\partial_{j)}\beta^k
 -\frac{2}{3}\widetilde{\gamma}_{ij}\partial_k\beta^k,
\label{eq:z4c_metric}\\
 \partial_\perp\widehat K={}&-D^iD_i\alpha
 +\alpha\left[
 \widetilde A_{ij}\widetilde A^{ij}
 +\frac{1}{3}(\widehat K+2\Theta)^2\right]
\nonumber\\
&+4\pi\alpha({\cal E}+{\cal S})
 +\alpha\kappa_1(1-\kappa_2)\Theta,
\label{eq:z4c_khat}\\
 \partial_\perp\widetilde A_{ij}={}&
 W^2\left[-D_iD_j\alpha
 +\alpha(R_{ij}-8\pi{\cal S}_{ij})\right]^{\rm TF}
\nonumber\\
&+\alpha\left[(\widehat K+2\Theta)\widetilde A_{ij}
 -2\widetilde A^k{}_i\widetilde A_{kj}\right]
\nonumber\\
&+2\widetilde A_{k(i}\partial_{j)}\beta^k
-\frac{2}{3}\widetilde A_{ij}\partial_k\beta^k,
\label{eq:z4c_aij}\\
 \partial_\perp\widetilde\Gamma^i={}&
 -2\widetilde A^{ij}\partial_j\alpha
 +2\alpha\widetilde\Gamma^i{}_{jk}\widetilde A^{jk}
 -\frac{6\alpha}{W}\widetilde A^{ij}\partial_jW
\nonumber\\
&-\frac{2\alpha}{3}\widetilde\gamma^{ij}
             \partial_j(2\widehat K+\Theta)
 -16\pi\alpha\widetilde\gamma^{ij}{\cal S}_j
\nonumber\\
&+\widetilde\gamma^{jk}\partial_j\partial_k\beta^i
 +\frac{1}{3}\widetilde\gamma^{ij}
                 \partial_j\partial_k\beta^k
\nonumber\\
&
 -(\widetilde\Gamma_{\rm d})^j\partial_j\beta^i
 +\frac{2}{3}(\widetilde\Gamma_{\rm d})^i\partial_j\beta^j
\nonumber\\
&-2\alpha\kappa_1
 \left[\widetilde\Gamma^i-(\widetilde\Gamma_{\rm d})^i\right],
\label{eq:z4c_gamma_evol}\\
 \partial_\perp\Theta={}&
 \frac{\alpha}{2}\left[
 R-\widetilde A_{ij}\widetilde A^{ij}
 +\frac{2}{3}(\widehat K+2\Theta)^2
 -16\pi{\cal E}\right]
 -\alpha\kappa_1(2+\kappa_2)\Theta .
\label{eq:z4c_theta}
\end{align}
Here $D_i$ is the covariant derivative compatible with $\gamma_{ij}$,
$R_{ij}$ and $R$ are its Ricci tensor and scalar, and TF denotes the
trace-free part with respect to $\gamma_{ij}$.  We adopt the damping
parameters used in the current implementation,
$\kappa_1=0.02$ and $\kappa_2=0$, in code units.  In the continuum
limit $\Theta=0$ and
$\widetilde{\Gamma}^i=(\widetilde{\Gamma}_{\rm d})^i$, and the above
system reduces to the BSSN evolution system.  At finite resolution,
the additional variables turn the Hamiltonian-constraint violation
into a propagating, damped mode instead of allowing a zero-speed mode
to remain near its point of origin.

The gauge conditions are an advective $1+\log$ slicing condition
written in terms of $\widehat K$ and a first-order Gamma-driver shift,
\begin{align}
 \partial_\perp\alpha &=-2\alpha\widehat K,
\label{eq:z4c_lapse}\\
 \partial_\perp\beta^i&=\widetilde\Gamma^i-\eta\beta^i,
\label{eq:z4c_shift}
\end{align}
where the implementation sets
$\eta=(3\,{\rm M}_{\odot})^{-1}$ in geometrized units.  After every
Runge--Kutta substep, the algebraic constraints
\begin{equation}
 \det(\widetilde\gamma_{ij})=1,\qquad
 \widetilde\gamma^{ij}\widetilde A_{ij}=0
\end{equation}
are imposed explicitly.
Following the prescription of \citet{Fujibayashi2020}, we apply a smooth radial window $\exp[-(r/r_\Theta)^2]$, with $r=\sqrt{x^2+y^2+z^2}$ being the radial distance and $r_\Theta=10^3\,{\rm km}$, for the non-advective right-hand side of equation~(\ref{eq:z4c_theta}).
This factor suppresses the Hamiltonian-constraint driving term in regions far from the coordinate centre.

\subsection{Radiation hydrodynamics}
\label{sec:radiation_hydrodynamics}

Apart from the change of the metric evolution system described above,
the general-relativistic radiation-hydrodynamics equations are the same
as those of \citet{Kuroda2016}.  For the hydrodynamic configuration used
in this work, the fluid stress-energy tensor is
\begin{equation}
 T_{\rm fluid}^{\mu\nu}=\rho h u^\mu u^\nu+P g^{\mu\nu},
\label{eq:fluid_stress_energy}
\end{equation}
where $\rho$, $P$, $u^\mu$, and
$h=1+e+P/\rho$ are the rest-mass density, pressure, fluid four-velocity,
and specific enthalpy, respectively, and $e$ is the specific internal
energy.  The total stress-energy tensor is
\begin{equation}
 T_{\rm total}^{\mu\nu}=T_{\rm fluid}^{\mu\nu}
 +\sum_s\int{\rm d}\varepsilon\,
 T_{(s,\varepsilon)}^{\mu\nu},
\label{eq:total_stress_energy}
\end{equation}
where $s$ labels the neutrino species and $\varepsilon$ is the neutrino
energy measured in the fluid-comoving frame.

We evolve the zeroth and first angular moments of the neutrino
distribution using the truncated-moment formalism
\citep{Thorne1981,Shibata2011}.  The spectral radiation stress-energy
tensor may be decomposed with respect to either an Eulerian observer or
the comoving observer:
\begin{align}
 T_{(\varepsilon)}^{\mu\nu}
 ={}&E_{(\varepsilon)}n^\mu n^\nu
 +F_{(\varepsilon)}^\mu n^\nu
 +F_{(\varepsilon)}^\nu n^\mu
 +P_{(\varepsilon)}^{\mu\nu},
\label{eq:rad_eulerian}\\
 ={}&J_{(\varepsilon)}u^\mu u^\nu
 +H_{(\varepsilon)}^\mu u^\nu
 +H_{(\varepsilon)}^\nu u^\mu
 +L_{(\varepsilon)}^{\mu\nu}.
\label{eq:rad_comoving}
\end{align}
The moments obey
$F^\mu n_\mu=P^{\mu\nu}n_\mu=0$ in the Eulerian frame and
$H^\mu u_\mu=L^{\mu\nu}u_\mu=0$ in the comoving frame.  We evolve
$E_{(\varepsilon)}$ and the covariant spatial components
$F_{(\varepsilon)i}$.  Suppressing the species label, their conservative
equations are
\begin{align}
 \partial_t(\sqrt{\gamma}E_{(\varepsilon)})={}&
 -\partial_i\left[
 \sqrt{\gamma}\left(\alpha F_{(\varepsilon)}^i
 -\beta^iE_{(\varepsilon)}\right)\right]
\nonumber\\
&-\sqrt{\gamma}\alpha\,
 \partial_\varepsilon\left(
 \varepsilon\widetilde M_{(\varepsilon)}^\mu n_\mu\right)
\nonumber\\
&+\sqrt{\gamma}\left[
 \alpha P_{(\varepsilon)}^{ij}K_{ij}
 -F_{(\varepsilon)}^i\partial_i\alpha
 -\alpha S_{(\varepsilon)}^\mu n_\mu\right],
\label{eq:radiation_energy}\\
 \partial_t(\sqrt{\gamma}F_{(\varepsilon)i})={}&
 -\partial_j\left[
 \sqrt{\gamma}\left(\alpha P_{(\varepsilon)i}{}^j
 -\beta^jF_{(\varepsilon)i}\right)\right]
\nonumber\\
&+\sqrt{\gamma}\alpha\,
 \partial_\varepsilon\left(
 \varepsilon\widetilde M_{(\varepsilon)}^\mu\gamma_{i\mu}\right)
\nonumber\\
&+\sqrt{\gamma}\left[
 -E_{(\varepsilon)}\partial_i\alpha
 +F_{(\varepsilon)j}\partial_i\beta^j
\right.\nonumber\\
&\left.\hspace{1.2em}
 +\frac{\alpha}{2}P_{(\varepsilon)}^{jk}
                         \partial_i\gamma_{jk}
 +\alpha S_{(\varepsilon)}^\mu\gamma_{i\mu}\right].
\label{eq:radiation_momentum}
\end{align}
Here $S_{(\varepsilon)}^\mu$ is the neutrino--matter interaction four-source, and
\begin{equation}
 \widetilde M_{(\varepsilon)}^\mu
 =M_{(\varepsilon)}^{\mu\alpha\beta}\nabla_\beta u_\alpha
\label{eq:energy_space_source}
\end{equation}
contains the gravitational-redshift and Doppler-shift terms.  The
rank-three moment $M^{\mu\alpha\beta}_{(\varepsilon)}$ is closed
consistently with the two-moment system.

The pressure tensor is supplied by an analytic M1 closure,
\begin{equation}
 P_{(\varepsilon)}^{ij}
 =\frac{3\chi_{(\varepsilon)}-1}{2}
  P_{\rm thin\,(\varepsilon)}^{ij}
 +\frac{3[1-\chi_{(\varepsilon)}]}{2}
  P_{\rm thick\,(\varepsilon)}^{ij},
\label{eq:m1_pressure}
\end{equation}
which interpolates between the free-streaming and diffusion limits.
The variable Eddington factor is given by the Minerbo closure
\citep{Minerbo1978},
\begin{align}
 \chi_{(\varepsilon)}
 &=\frac{5+6f_{(\varepsilon)}^2
 -2f_{(\varepsilon)}^3+6f_{(\varepsilon)}^4}{15},
\label{eq:minerbo}\\
 f_{(\varepsilon)}^2
 &\equiv
 \frac{h_{\mu\nu}H_{(\varepsilon)}^\mu
 H_{(\varepsilon)}^\nu}{J_{(\varepsilon)}^2},\qquad
 h_{\mu\nu}=g_{\mu\nu}+u_\mu u_\nu .
\label{eq:comoving_flux_factor}
\end{align}
Because $J_{(\varepsilon)}$ and $H_{(\varepsilon)}^\mu$ themselves
depend on $P_{(\varepsilon)}^{ij}$ through the transformation between
the Eulerian and comoving frames, equations~(\ref{eq:m1_pressure})--(\ref{eq:comoving_flux_factor}) are solved iteratively.  This construction
ensures $\chi\rightarrow1/3$ in the optically thick limit and
$\chi\rightarrow1$ in the free-streaming limit.
In the diffusion limit, we evaluate the numerical fluxes for both the zeroth ($E$) and first ($F_i$) moments using a control parameter originally proposed by \citet{Audit2002} and subsequently adopted in many moment-based transport codes \citep{OConnorOtt2013,Kuroda2016}, thereby enforcing the correct asymptotic diffusion flux.

The fluid equations are also written in conservative form.  Defining
the fluid Lorentz factor
$W_{\rm L}\equiv\alpha u^t$, the coordinate velocity
$v^i\equiv u^i/u^t$, and
\begin{equation}
D=\rho W_{\rm L},\quad
 S_i=\rho hW_{\rm L}u_i,\quad
 S_0=\rho hW_{\rm L}^2-P,\quad
 \tau=S_0-D,
\label{eq:hydro_variables}
\end{equation}
we solve
\begin{eqnarray}
&& \partial_t(\sqrt{\gamma}D)+\partial_i(\sqrt{\gamma}D v^i)=0,
\label{eq:mass_conservation}\\
&& \partial_t(\sqrt{\gamma}S_i)
 +\partial_j\left[\sqrt{\gamma}
 (S_iv^j+\alpha P\delta_i{}^j)\right]=
\nonumber\\
&&+\sqrt{\gamma}
 (-S_0\partial_i\alpha+S_k\partial_i\beta^k)
+\frac{\alpha\sqrt{\gamma}}{2}
 S^{jk}\partial_i\gamma_{jk}
 -\alpha\sqrt{\gamma}\int{\rm d}\varepsilon\,
 S_{(\varepsilon)}^\mu\gamma_{i\mu},\nonumber \\
\label{eq:fluid_momentum}\\
&& \partial_t(\sqrt{\gamma}\tau)
 +\partial_i\left\{\sqrt{\gamma} [\tau v^i+P(v^i+\beta^i)]\right\}=\nonumber \\
&&+\sqrt{\gamma}
 (\alpha S^{ij}K_{ij}-S_iD^i\alpha)
+\alpha\sqrt{\gamma}\int{\rm d}\varepsilon\,
 S_{(\varepsilon)}^\mu n_\mu .
\label{eq:fluid_energy}
\end{eqnarray}
The opposite signs of the interaction terms in
equations~(\ref{eq:radiation_energy})--(\ref{eq:radiation_momentum}) and
(\ref{eq:fluid_momentum})--(\ref{eq:fluid_energy}) enforce exchange,
rather than creation or destruction, of the total energy and momentum.
The electron fraction is evolved according to
\begin{align}
 \partial_t(\sqrt{\gamma}D Y_e)+\partial_i(\sqrt{\gamma} D Y_ev^i)
 ={}&\sqrt{\gamma}\alpha m_{\rm u}
 \int\frac{{\rm d}\varepsilon}{\varepsilon}
 \left(S_{(\nu_e,\varepsilon)}^\mu
 -S_{(\bar\nu_e,\varepsilon)}^\mu\right)u_\mu ,
\label{eq:electron_fraction}
\end{align}
and the corresponding total lepton fraction is monitored to maintain
the conservative coupling between matter, electron neutrinos, and
electron antineutrinos.

The default three-species configuration evolves $\nu_e$, $\bar\nu_e$,
and a grouped heavy-lepton species $\nu_x$; the latter represents
$\nu_\mu$, $\bar\nu_\mu$, $\nu_\tau$, and $\bar\nu_\tau$ and therefore
enters energy-integrated quantities with a multiplicity of four.  A
six-species build is also supported.  We discretize the comoving-energy
domain into $N_\varepsilon$ groups, whose boundaries are denoted by
$\varepsilon_{n-1/2}$ and $\varepsilon_{n+1/2}$
($n=1,\ldots,N_\varepsilon$), with $\varepsilon_{1/2}=0$.
The energy interval and group spacing are configurable and are specified for each calculation.
The collision term includes charged-current emission and absorption on free nucleons and nuclei, isoenergetic scattering on nucleons and nuclei, inelastic neutrino--electron scattering, electron--positron pair processes, and nucleon--nucleon bremsstrahlung, as in \citet{Kotake2018,Kuroda2016}.
For each neutrino species and energy group, the interaction four-source term can be written as
\begin{align}
 S_{(\varepsilon)}^\mu={}&\kappa_{\rm a}(J^{\rm eq}-J)u^\mu
 -(\kappa_{\rm a}+\kappa_{\rm s})H^\mu \nonumber\\
 &+a_0(-J u^\mu-H^\mu)
 +c_0\left[(4\pi\varepsilon^3-J)u^\mu-H^\mu\right].
\label{eq:collision_source}
\end{align}
Here, $J$ and $H^\mu$ are the comoving-frame zeroth and first radiation moments, respectively, with $H^\mu u_\mu=0$; $J^{\rm eq}$ is the equilibrium moment constructed from the emissivity and absorption opacity.
The coefficients $\kappa_{\rm a}$ and $\kappa_{\rm s}$ correspond to the absorption and isoenergetic-scattering coefficients, while $a_0$ and $c_0$ collect the zeroth-angular-moment energy-coupling terms from neutrino--electron scattering, pair processes, and nucleon--nucleon bremsstrahlung.
The factor $4\pi\varepsilon^3$ is the fully occupied phase-space moment in the code normalization.
Spatial transport and geometrical source terms are operator split from the stiff neutrino--matter coupling, for which the radiation moments and thermodynamic variables are updated together by an implicit iteration, following the same procedure as described in \citet{Kuroda2016}.

\section{Kokkos implementation}
\label{sec:implementation}

\subsection{Design and portability strategy}
\label{sec:kokkos_design}

The accelerator version is written in C++17 and uses Kokkos to separate the expression of on-node parallelism and memory placement from the execution back-end \citep{Edwards2014,Trott2022}.
We preserved the algorithmic structure of the original Fortran code wherever possible: the finite-difference stencils, reconstruction, Riemann solver, M1 closure, neutrino interaction rates, equation-of-state (EOS) interpolation, and time-integration sequence are common to the two implementations.
The principal change is therefore associated with the organization of data and work required for accelerator execution.
This correspondence also permits individual stages of the Kokkos code to be compared directly with their Fortran counterparts.

The implementation uses \texttt{Kokkos::DefaultExecutionSpace} and its associated memory
space for all state arrays and computational kernels.  
Consequently, the physics source does not contain device-specific kernel-launch code.
The build used for the MI300A calculations selects the Kokkos HIP back-end and the AMD \texttt{gfx942} target; serial and OpenMP host back-ends are also enabled in the build system.
The executable is linked to MPI, and the present HIP build uses relocatable device code.
The exact compiler, Kokkos, ROCm, and MPI versions used for the measurements are reported together with the machine configurations in Section~\ref{sec:performance}.

Compile-time parameters specify the numbers of spatial subblocks $n_x$, $n_y$, and $n_z$ per FMR level in the three Cartesian directions, the number $n_{\rm cell}$ of active cells per direction within each subblock, the number $n_{\rm ghost}$ of ghost cells on each side, the number of refinement levels $L_{\rm FMR}$, the number of neutrino species $N_\nu$, and the number of neutrino-energy groups $N_\varepsilon$.
The three-dimensional Cartesian computational domain contains a nested hierarchy of $L_{\rm FMR}$ fixed-mesh-refinement levels, and the domain covered by each level is partitioned into $n_xn_yn_z$ spatial subblocks.
The calculation therefore uses a total of $n_xn_yn_z$ MPI ranks, with each rank assigned to one spatial subblock and owning the corresponding subblock at every refinement level from $l=0$ to $l=L_{\rm FMR}-1$.
Each spatial subblock is further discretized into $n_{\rm cell}^3$ active cells, or $n_{\rm cell}$ cells along each Cartesian direction, together with $n_{\rm ghost}$ ghost cells on each side.
In the current implementation, we set $n_x=n_y=n_z$ and define the total number of active cells per Cartesian direction on each FMR level by $N_{\rm cell}\equiv n_xn_{\rm cell}$.
Thus, every FMR level contains $n_x^3$ spatial subblocks and is resolved by $N_{\rm cell}$ active cells along each Cartesian direction.

Keeping these dimensions, including $N_\nu$ and $N_\varepsilon$, at compile time allows the compiler to determine the sizes of the cell-local work arrays used by the radiation and microphysics kernels.

\subsection{Data layout and parallel decomposition}
\label{sec:kokkos_data}

The evolved and auxiliary fields are stored in one-dimensional
\texttt{Kokkos::View} objects of double-precision values.  A logical
multidimensional index is converted by a device-callable inline
function to
\begin{equation}
 {\cal I}(l,k,j,i,q)
 = \left[\left(\left(l\tilde{n}_z+k\right)\tilde{n}_y+j\right)\tilde{n}_x+i\right]N_q+q .
\label{eq:flattened_index}
\end{equation}
Here $l$ is the refinement-level index, $(i,j,k)$ are the corresponding cell indices, and $q$ denotes the component stored at a cell.
The quantities $\tilde{n}_x$, $\tilde{n}_y$, and $\tilde{n}_z$ are the ghost-inclusive local array extents; for the cubic blocks used here, they satisfy $\tilde{n}_x=\tilde{n}_y=\tilde{n}_z=n_{\rm cell}+2n_{\rm ghost}$.
The refinement-level index runs from $l=0$ to $l=L_{\rm FMR}-1$.
Because $\tilde{n}_x=\tilde{n}_y=\tilde{n}_z$ for the cubic blocks, each of the local cell indices $i$, $j$, and $k$ runs from 0 to $\tilde{n}_x-1$.
These tilded quantities are distinct from $n_x$, $n_y$, and $n_z$ defined above, which denote the numbers of spatial subblocks in the three Cartesian directions.
Field components belonging to the same cell are therefore contiguous in memory.
For the spectral radiation fields, $q$ is further
factorized into species, energy, and moment indices.
This layout gives each work item contiguous access to the hydrodynamic variables or radiation moments that it uses, while retaining a simple common indexing scheme for the CPU and accelerator back-ends.
Separate \texttt{View} objects hold the current and beginning-of-step states, reconstructed left/right states, directional fluxes, gravitational source terms, redshift and Doppler terms, M1 closure quantities, opacities, and EOS interpolation results.

At runtime, MPI provides the inter-block decomposition.
Spatial parallelism within the rank is exposed through Kokkos.
In the commonly used kernel pattern, the level and three-dimensional cell indices are collapsed into a single \texttt{RangePolicy}; kernels that naturally contain additional independent work also include the spatial direction or the species--energy pair in this flattened index.
This construction supplies many more work items than there are cells in one $n_{\rm cell}^3$ subblock and avoids a serial loop over the $N_\varepsilon$ energy groups.

Ghost exchange is also performed with device-resident Views.
A Kokkos kernel packs all faces at a given level into contiguous send buffers, after which their \texttt{data()} pointers are passed directly to a GPU-aware MPI implementation.
Persistent point-to-point MPI requests are used for repeated same-level exchanges.
Receive completions are processed with \texttt{MPI\_Waitany}, so that the unpack kernel for one direction can be launched while other directions remain in flight.
Copies for local or symmetry-related neighbours use
\texttt{Kokkos::subview} and \texttt{Kokkos::deep\_copy} and therefore remain in the device memory space.

Coarse--fine boundaries require additional communication paths.
The restriction and prolongation operators are Kokkos kernels, and their MPI buffers are likewise device resident.
The hydrodynamics and transport update additionally applies flux restriction at FMR interfaces: fine-grid face fluxes are packed and communicated, and the corresponding coarse-grid flux is replaced by the appropriately restricted value before the conservative update.
Hence no host staging is required by the regular time-stepping path, provided that the MPI library supports device pointers.

\subsection{Accelerated kernels}
\label{sec:kokkos_kernels}

All principal stages of a time step execute as Kokkos kernels.
For the spacetime sector, these comprise evaluation of the Z4c right-hand side, the Runge--Kutta update, and enforcement of the algebraic conformal constraints.
The metric variables are advanced with four Runge--Kutta stages, with a ghost-zone exchange after each stage.

The radiation-hydrodynamics sector is advanced in two stages.
Each stage performs, in order, piecewise-parabolic reconstruction, numerical-flux evaluation for the fluid and multi-energy neutrino radiation moments, refluxing (i.e. flux correction at FMR boundaries), evaluation of geometrical source terms, gravitational-redshift and Doppler terms in energy space, the conservative update, conservative-to-primitive recovery, the neutrino--matter interaction update, and ghost exchange.
The M1 closure and opacity kernels are also executed on the accelerator and are refreshed when required by the transport, interaction, ghost-zone, or diagnostic operations.

The EOS and weak-interaction data are represented by read-only device-resident tables.
The neutrino source kernel assigns one active cell to a work item and performs the coupled, cell-local implicit update of matter and all spectral moments on the device.
Its temporary vectors and small matrices use fixed-size private arrays, implemented primarily with \texttt{Kokkos::Array}; the same approach is used for metric tensors, primitive-recovery workspaces, and the M1 transformations.
This eliminates dynamic allocation inside kernels and makes the data required for a single cell available to the compiler.

Most kernels use \texttt{Kokkos::parallel\_for} with a one-dimensional \texttt{RangePolicy}.
Device-callable numerical routines are marked with \texttt{KOKKOS\_INLINE\_FUNCTION}, and kernel bodies are expressed with \texttt{KOKKOS\_LAMBDA}.
Kernels submitted to the same execution space retain their data dependencies without a host synchronization after every launch.
Explicit \texttt{Kokkos::fence} calls are introduced at visibility boundaries: before MPI consumes a packed device buffer, after receives whose contents are needed by subsequent work, before host-side diagnostics or output, and between dependent FMR communication operations.
This organization keeps the computational state resident on the accelerator throughout the complete update while making the synchronization points visible in profiles.

\section{Code validation}
\label{sec:verification}
This section presents a variety of validation tests.
Sections~\ref{sec:Linear wave test in 2D} and \ref{sec:Shock tube test} are performed in two and one spatial dimensions, respectively.
All remaining tests are performed in three dimensions, but we evolve only one octant ($x,y,z\ge0$) and impose octant symmetry.

\subsection{Linear wave test in 2D}
\label{sec:Linear wave test in 2D}
\begin{figure}
\begin{center}
\includegraphics[angle=0.,width=\columnwidth]{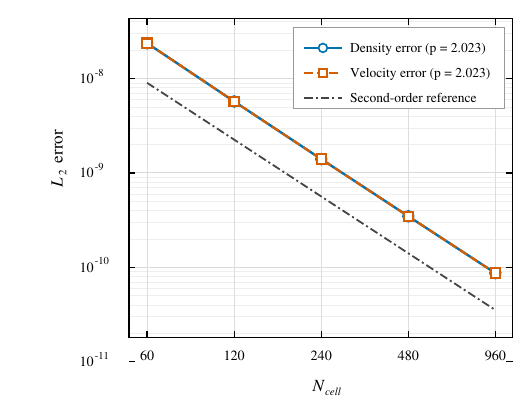}
\caption{Resolution dependence of the $L_2$ errors in density (circles) and
velocity (squares) for the two-dimensional adiabatic hydrodynamic linear-wave
test.  The dash-dotted line denotes the expected second-order scaling.  Fits
to the numerical results give convergence indices of $2.023$ for both
quantities, demonstrating second-order convergence of the multidimensional
hydrodynamics update with periodic boundaries and HLLC fluxes.
\label{fig:linear_wave_convergence}}
\end{center}
\end{figure}
We first test the multidimensional hydrodynamics solver using the 2D adiabatic linear-wave problem.
A small-amplitude ($10^{-6}$) eigenmode is superposed on a uniform, non-magnetized background and propagated obliquely ($\theta=\tan^{-1}(0.5)$) across a square domain of $(x,y)\in[0,1]$ covered by $(N_{\rm cell})^2$ cells.
Periodic boundary conditions are imposed in both directions.
We measure the $L_2$ norm of the deviation from the exact solution for the density and velocity components after one wave-crossing time.
We apply PPM reconstruction and use the HLLC numerical flux \citep{Mignone2005}, although the current test is entirely in the Newtonian limit.
To check the numerical convergence, we consider five different resolutions $N_{\rm cell}=60$, $120$, $240$, $480$, and $960$.
Figure~\ref{fig:linear_wave_convergence} shows that the code achieves the expected second-order convergence.

\subsection{Shock tube test}
\label{sec:Shock tube test}
\begin{figure}
\begin{center}
\includegraphics[angle=0.,width=\columnwidth]{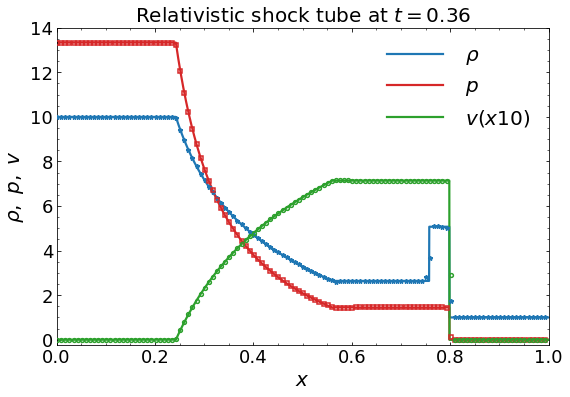}
\caption{Relativistic shock-tube solution at $t=0.36$.  The rest-mass density
$\rho$, pressure $p$, and ten times the normal three-velocity $v_x$ are shown.
Continuous curves give the exact special-relativistic Riemann solution and
symbols show the numerical result.  The numerical profiles reproduce the
smooth rarefaction and the positions and intermediate states of the contact
discontinuity and leading shock.
\label{fig:ShockTube}}
\end{center}
\end{figure}
We next consider the mildly relativistic blast-wave problem commonly used to test special-relativistic hydrodynamics schemes \citep{MartiMueller2003}.
The initial discontinuity at $x=0.5$ separates the states
\begin{equation}
 (\rho,p,v_x)_L=(10,40/3,0),\qquad
 (\rho,p,v_x)_R=(1,2\times10^{-6}/3,0),
\end{equation}
with an ideal-gas adiabatic index $\Gamma=5/3$.
The decay of the discontinuity produces a left-going rarefaction, a contact discontinuity, and a right-going shock.
This test simultaneously probes primitive-variable recovery, relativistic characteristic speeds, the special relativistic HLLC Riemann solver, and shock capturing over a large pressure contrast.
In this test, we use piecewise-linear reconstruction (PLM) with a resolution of $N_{\rm cell}=600$.
Figure~\ref{fig:ShockTube} plots the exact solution (lines) and our numerical results (symbols) for the density (blue), pressure (red), and three-velocity (green, multiplied by ten) at $t=0.36$.
The numerical results reproduce the exact solution well without spurious oscillations.

\subsection{TOV star}
\label{sec:TOV star}

We next test the coupled hydrodynamics and dynamical-spacetime sectors by evolving a stable, non-rotating relativistic star in hydrostatic equilibrium.
The initial model is constructed by integrating the Tolman--Oppenheimer--Volkoff equations.
We use the cold polytropic equation of state $P=K\rho^\Gamma$ with \(\Gamma=2\), \(K=2.17\times10^{-4}\), and central rest-mass density \(\rho_{\rm c}=6.0\times10^2\) in code units (\(G=c=1\)).
For the adopted code length scale ($10^3$\,km), the central density in cgs units is \(8.08\times10^{14}\ {\rm g\,cm^{-3}}\).
The resulting equilibrium model has a gravitational mass of \(1.405\,{\rm M_\odot}\), an areal radius of \(14.07\) km, and compactness \(M/R=0.147\).

\begin{figure}
\begin{center}
\includegraphics[angle=0.,width=\columnwidth]{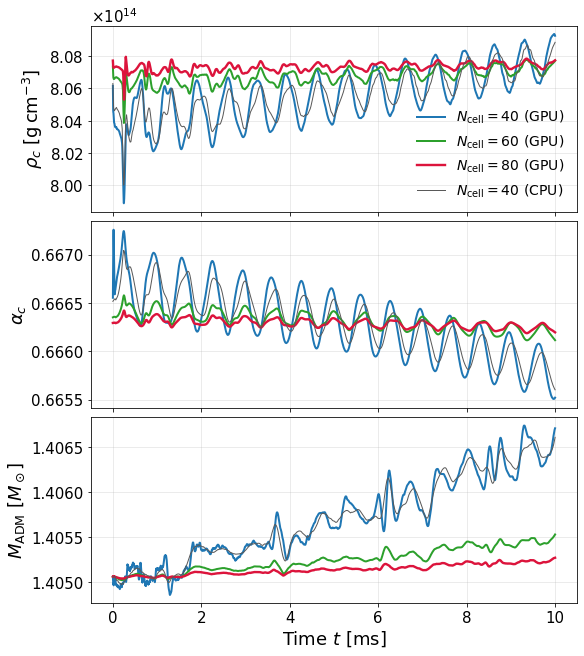}
\caption{Evolution of the non-rotating TOV star at three finest-level grid
spacings.  From top to bottom, the panels show the central rest-mass density
\(\rho_{\rm c}\), central lapse \(\alpha_{\rm c}\), and ADM mass
\(M_{\rm ADM}\).
The thick coloured curves show the GPU results at the three resolutions, whereas the thin grey curve shows the SAKURA CPU result at $N_{\rm cell}=40$.
The finite-volume representation of the equilibrium model
excites small radial oscillations.  Their amplitudes, together with the
secular drift of the ADM mass, decrease as the grid is refined from
\(\Delta x=390\) to \(195\) m.
\label{fig:TOVstar}}
\end{center}
\end{figure}

We evolve the model for \(10\) ms at three resolutions, \(\Delta x=390\), \(260\), and \(195\) m on the finest refinement level, corresponding to \(N_{\rm cell}=40\), 60, and 80 cells per FMR level, respectively, with $L_{\rm FMR}=6$ refinement levels in a computational domain of $(500\,{\rm km})^3$.
Mapping the spherical equilibrium solution to a finite Cartesian grid
introduces truncation-level perturbations that excite radial pulsations.
Figure~\ref{fig:TOVstar} shows that their amplitudes in both the central
density and central lapse decrease systematically with increasing
resolution.  The secular change in the ADM mass is likewise reduced: over
the simulated interval it remains below approximately \(0.1\) per cent at
the lowest resolution and \(0.02\) per cent at the highest resolution.
The bounded oscillations and their resolution dependence demonstrate that
the Kokkos implementation preserves the relativistic stellar equilibrium
and consistently couples the hydrodynamic and BSSN--Z4c updates.

For the CPU comparison in Figure~\ref{fig:TOVstar}, we used one node of the SAKURA cluster at our institute.\footnote{\href{https://docs.mpcdf.mpg.de/doc/computing/clusters/systems/Gravitational_Physics.html}{MPCDF SAKURA}.}
Each SAKURA compute node contains two Intel Xeon Gold 6248 processors at $2.5$~GHz, with 20 physical cores per socket and 376~GB of memory.
On this system, we additionally compiled \codeName~for CPU execution and evolved the $N_{\rm cell}=40$ model.
Although the CPU and GPU calculations were performed on different computing systems using different compilers, their evolutions of the central density, central lapse, and ADM mass agree closely throughout the simulated interval.
The comparison therefore demonstrates that \codeName~produces numerically consistent results on CPU and GPU architectures.

\subsection{Diffusion test}
\label{sec:Diffusion test}

\begin{figure}
\begin{center}
\includegraphics[angle=0.,width=\columnwidth]{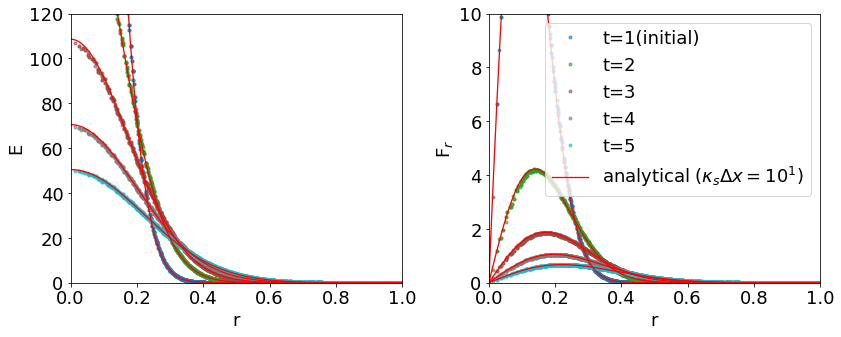}
\includegraphics[angle=0.,width=\columnwidth]{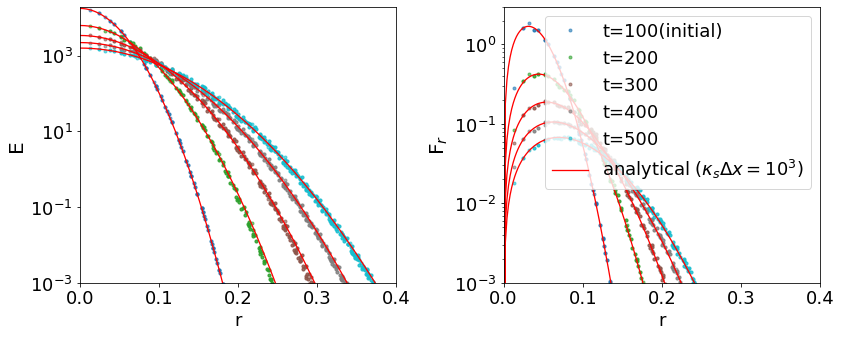}
\caption{Diffusion of a spherical radiation pulse in a static, purely
scattering medium.  The left and right columns show the radiation energy
density $E$ and radial radiation flux $F^r$, respectively.  The upper row
uses $\kappa_s\Delta x=10$, whereas the lower row uses $\kappa_s \Delta x=10^3$ and is shown
with logarithmic ordinates to expose the pulse tails.  Numerical profiles at the
times indicated in the legends are compared with the analytic diffusion
solution.  Their agreement demonstrates that both radiation moments evolve
correctly in the optically thick limit, including the high-opacity case.
\label{fig:DiffTest}}
\end{center}
\end{figure}

The optically thick limit of the radiation-moment solver is verified using
the diffusion-wave test \citep{Pons2000}.
A point-like radiation source initially placed at the origin diffuses through a static
medium with zero absorption and constant scattering opacity $\kappa_s$.
The analytic solution for the radiation energy density and radial flux at a radius $r$ and time $t$ is given by
\begin{equation}
 E(r,t)=\left(\frac{\kappa_s}{t}\right)^{3/2}
 \exp\left(-\frac{3\kappa_s r^2}{4t}\right),\qquad
 F^r(r,t)=\frac{r}{2t}E(r,t).
\end{equation}
We consider two dimensionless opacities, $\kappa_s\Delta x=10$ and $10^3$, where $\Delta x$ denotes the grid width, thereby testing both a moderately diffusive case and a strongly opaque regime in which the transport flux must approach its asymptotic diffusion limit.
Figure~\ref{fig:DiffTest} shows good agreement between the numerical results (points) and the analytic solutions (lines).

\begin{figure}[ht]
\begin{center}
\includegraphics[angle=0.,width=\columnwidth]{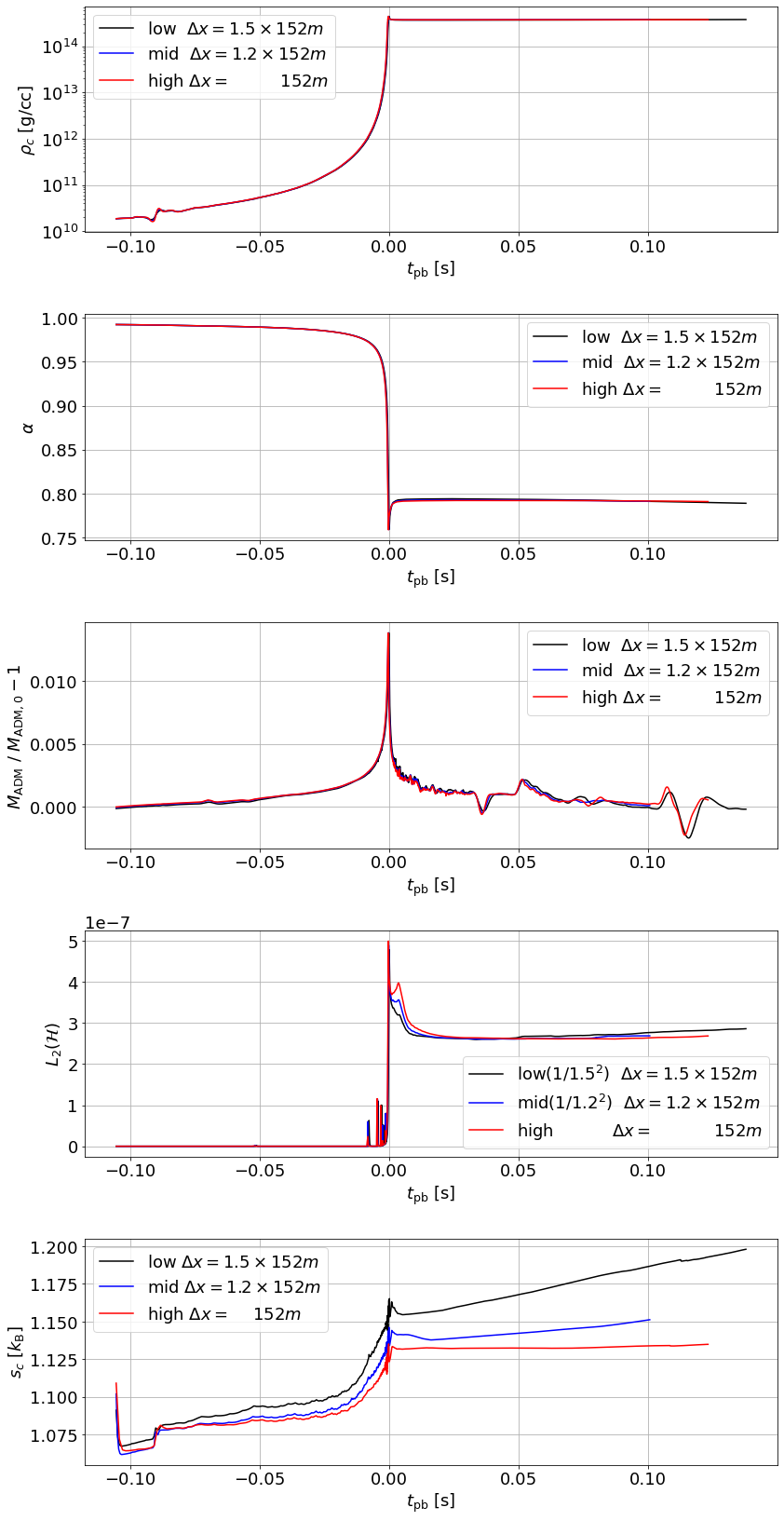}
\caption{Adiabatic collapse and core bounce of the $50\,M_\odot$ progenitor
at three spatial resolutions.  From top to bottom, the panels show the
central rest-mass density $\rho_c$, central lapse $\alpha_c$, fractional
deviation of the ADM mass from its initial value
$M_{\rm ADM}/M_{\rm ADM,0}-1$, $L_2$ norm of the Hamiltonian-constraint
violation, and central entropy per baryon.  Time is measured relative to core
bounce, $t_{\rm pb}=0$.  The close agreement of the three resolutions in the
collapse and bounce dynamics, together with the controlled ADM-mass error
and Hamiltonian-constraint violation, verifies the coupled GR-hydrodynamics
evolution.  Because the collapse is adiabatic, the resolution dependence of
the central entropy provides a sensitive measure of numerical entropy
production during bounce.
\label{fig:AdiabaticCollapse}}
\end{center}
\end{figure}

\subsection{Adiabatic collapse of a massive star}
\label{sec:Adiabatic Collapse of Massive Star}

We also test the coupled hydrodynamics and dynamical-spacetime sectors by evolving the adiabatic gravitational collapse of the non-rotating $50\,M_\odot$ progenitor of \citet{UmedaNomoto2008}.
Neutrino transport and weak-interaction source terms are omitted in this controlled calculation, so that the collapse, core bounce, and subsequent prompt-shock propagation provide direct tests of the relativistic fluid-gravity coupling.
For the present test, we use the SFHo nuclear relativistic mean-field equation of state (EOS) of \citet{SFH}, including contributions from electrons, positrons, and photons.
Three different resolutions are considered: $N_{\rm cell}=96$, $80(=96\times1.2^{-1})$, and $64(=96\times1.5^{-1})$, with a fixed number of refinement levels, $L_{\rm FMR}=10$, and a numerical box size of $L=1.5\times10^9$\,cm, resulting in central resolutions of $\Delta x=152$, $182$, and $228$\,m, respectively.
In this test, we monitor both central quantities and global or local measures of numerical accuracy to verify the code in a practical test problem and assess its numerical convergence.

Figure~\ref{fig:AdiabaticCollapse} shows, from top to bottom, the central rest-mass density, central lapse function, relative deviation of the Arnowitt--Deser--Misner (ADM) mass from its initial value, the $L_2$ norm of the Hamiltonian-constraint violation ($L_2(\mathcal H)$), and central entropy.
For the medium- and low-resolution models, $L_2(\mathcal H)$ is multiplied by $1.2^{-2}$ and $1.5^{-2}$, respectively; the scaled curves should therefore approximately overlap if the test achieves second-order convergence.

This progenitor model is known for its relatively high compactness ($\xi_{2.5}=0.19$) \citep[see][for the definition of $\xi_{2.5}$]{OConnorOtt2011}, enabling a rapid bounce at around $105$\,ms after the initiation of collapse.
All evolutions behave as expected in the adiabatic case.
The central density and lapse remain nearly constant after bounce without any signature of proto-neutron star (PNS) contraction, because of the absence of deleptonization and neutrino cooling, and the resulting prompt explosion.
The ADM mass is conserved within $\approx0.2$\,\%, which is expected in the adiabatic case, where the energy loss (e.g., via neutrinos) from the system is absent.
The evolution of $L_2(\mathcal H)$ indicates that the code achieves the designed second-order convergence of the global Hamiltonian-constraint error.
The central entropy deviates by $\approx4$--$9$\,per cent, with the highest-resolution model exhibiting the smallest deviation and an approximately constant value after bounce.
Although the central $Y_e$ evolution is not shown, $Y_e$ remains close to its initial value throughout the adiabatic collapse and post-bounce phase, with deviations below $\approx0.4$\,per cent.
It is known that the prompt shock launched at the bounce does not stall in the absence of neutrino cooling.
The adiabatic test reproduces this behaviour, with no clear shock stagnation, as expected.
This test therefore provides an additional verification of the shock-capturing scheme in a more realistic setup.

\section{Core-collapse supernova application}
\label{sec:application}

To demonstrate that \codeName~can carry out an end-to-end simulation with the physics required for a realistic core-collapse supernova application, we simulate the collapse, core bounce, and early post-bounce evolution of a massive-star progenitor.
We adopt the reference setup of the spherical code-comparison study by \citet{OConnor2018}, which enables a direct comparison of the hydrodynamic and neutrino signals with six independently developed supernova codes.
We use the same non-rotating, solar-metallicity $20\,M_\odot$ progenitor model of \citet{WH2007} and the SFHo \citep{SFH} nuclear equation of state as in that study.
We also adopt the basic neutrino-opacity set prescribed in \citet{OConnor2018}.
Specifically, scattering and absorption on free nucleons follow the rates of \citet{Bruenn1985}, supplemented by the weak-magnetism and recoil corrections of \citet{Horowitz2002}.
For charged-current absorption on free nucleons, no nucleon mean-field potentials are included beyond the neutron--proton rest-mass difference, consistently with the comparison setup.
Coherent scattering on heavy nuclei uses the \citet{Bruenn1985} rate, with ion--ion correlation corrections from \citet{Horowitz1997} and nuclear-form-factor corrections following \citet{BruennMezzacappa1997} and \citet{RamppJanka2002}.
Electron-neutrino absorption on nuclei and inelastic neutrino--electron scattering are likewise evaluated with the rates of \citet{Bruenn1985}.
For thermal pair production, electron--positron annihilation follows \citet{Bruenn1985}, whereas nucleon--nucleon bremsstrahlung follows \citet{HannestadRaffelt1998}.
The transport distinguishes $\nu_e$, $\bar{\nu}_e$, and a grouped heavy-lepton species $\nu_x=\{\nu_\mu,\bar{\nu}_\mu,\nu_\tau,\bar{\nu}_\tau\}$.
As in \citet{OConnor2018}, scattering and absorption on the SFHo light clusters ($^2{\rm H}$, $^3{\rm H}$, and $^3{\rm He}$) are neglected, and those clusters are not reassigned to free nucleons, alpha particles, or heavy nuclei.
The multi-energy M1 transport uses $N_\varepsilon=12$ energy groups for each of the three neutrino species.

The spatial decomposition uses $n_x=n_y=n_z=2$ subblocks per FMR level and $n_{\rm cell}=20$ active cells per direction within each subblock, giving $N_{\rm cell}=n_xn_{\rm cell}=40$ active cells per direction across every FMR level.
With $L_{\rm FMR}=11$ and a numerical box size of $L=1.5\times10^9$\,cm, this setup gives a central resolution of $\Delta x=336$\,m.
We should note that the resulting level-wide resolution of $N_{\rm cell}=40$ used in this test is not sufficiently high for practical CCSN simulations, although the central resolution of $\Delta x=336$\,m may be adequate.
With this configuration, a typical standing-shock radius at $\sim100$\,km is resolved with $\Delta x=2.7$\,km, corresponding to an effective angular resolution of $\approx4.9^\circ$.
Recent three-dimensional CCSN production simulations commonly employ an angular resolution of approximately $1.4^\circ$ \citep{PowellEtAl2023,BurrowsEtAl2024}, while a recent dedicated resolution study spans $0.7^\circ$--$2.8^\circ$ \citep{VarmaMueller2026}.
Thus, our effective angular resolution of $4.9^\circ$ is approximately 3.5 times coarser than the resolution typically adopted in recent simulations.
This may introduce larger numerical perturbations and result in a non-negligible deviation between this test and other studies.

Given the available computational resources, we restrict the present simulation to approximately the first $100$~ms after core bounce.
This interval nevertheless covers the pre-bounce collapse dynamics and the early post-bounce phase, before multidimensional convection has become fully developed.
Because we use the same progenitor and the same SFHo nuclear equation of state as in the reference simulations, this interval permits a direct and detailed comparison of both the basic collapse dynamics and the early neutrino profiles with the one-dimensional, spherically symmetric results of \citet{OConnor2018}.
The primary purpose of this test is to verify that the multi-energy neutrino-transport terms and the neutrino--matter interaction source terms are incorporated consistently in \codeName, rather than to follow the later nonlinear multidimensional evolution.

\begin{figure}
  \centering
\includegraphics[width=\columnwidth]{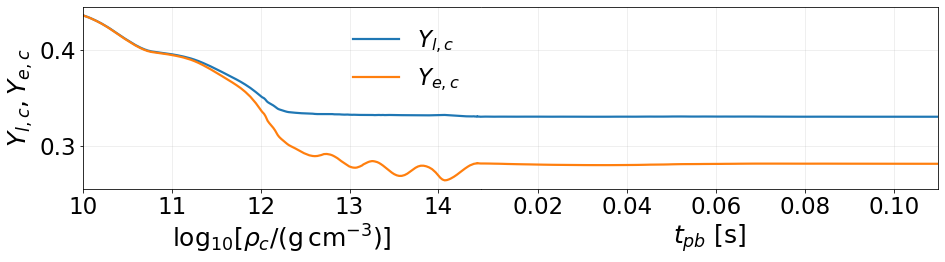}
  \caption{Evolution of the central total lepton fraction $Y_{l,c}$ and electron fraction \(Y_{e,c}\) in the core-collapse supernova application.
The left part shows the collapse phase as a function of the central rest-mass density \(\rho_c\), while the right part shows the post-bounce evolution as a function of \(t_{\rm pb}\).
The vertical dashed line marks core bounce.}
  \label{fig:central_lepton}
\end{figure}
Core bounce occurs at $291.1$~ms after the start of the simulation.
For the same progenitor and comparison setup, \citet{OConnor2018} reported collapse times of approximately 275~ms for 3DnSNe-IDSA, 419~ms for AGILE-BOLTZTRAN, 299.5~ms for FORNAX, 298.2~ms for GR1D, and 297.8~ms for PROMETHEUS-VERTEX.
The ANRe-M1 value lies within this reported range and is particularly close to the 297.8--299.5~ms cluster of FORNAX, GR1D, and PROMETHEUS-VERTEX, differing by only 6.8--8.5~ms (2.3--2.8 per cent).
This agreement provides an additional check that the collapse dynamics and the timing of core bounce are reproduced consistently.
As a next diagnostic of the core-collapse simulation, Figure~\ref{fig:central_lepton} shows the central electron fraction $Y_{e,c}$ and total lepton fraction $Y_{l,c}$ during collapse as functions of the central rest-mass density (left part), and after core bounce as functions of the post-bounce time (right part).
At low density the two quantities closely track one another as electron capture reduces the electron fraction, while neutrinos escape nearly freely from the core.
Once neutrinos become trapped at $\rho\gtrsim10^{12}$\,g\,cm$^{-3}$, \(Y_{l,c}\) separates from \(Y_{e,c}\) and remains approximately constant near 0.33 through the final phase of collapse, whereas \(Y_{e,c}\) continues to decrease to about 0.28 at bounce.
This evolution exhibits the deleptonization and neutrino trapping expected in a realistic core-collapse supernova.
After bounce, both central values remain nearly constant.

\begin{figure*}
\centering
\begin{minipage}[t]{0.49\textwidth}
\centering
\includegraphics[width=\linewidth]{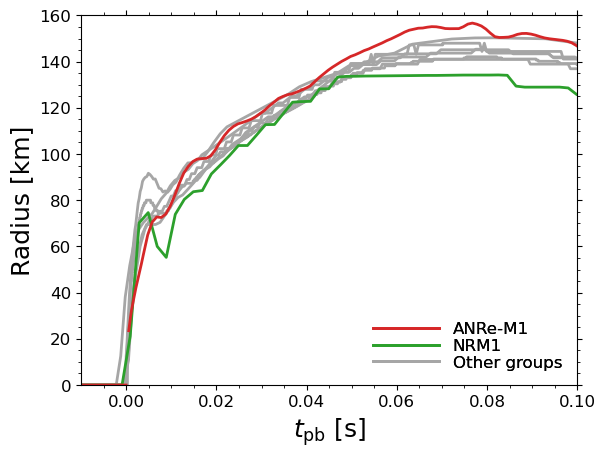}
\end{minipage}
\hfill
\begin{minipage}[t]{0.49\textwidth}
\centering
\includegraphics[width=\linewidth]{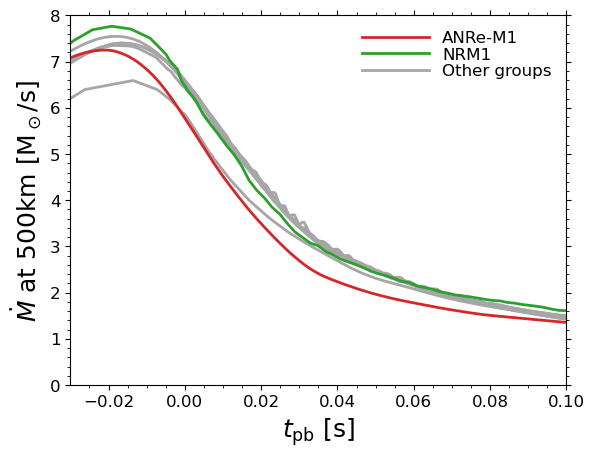}
\end{minipage}
\caption{Hydrodynamic comparison for \codeName (red), the CPU-based NRM1 code (green), and the six other code groups participating in the comparison of \citet{OConnor2018}. Left: shock radius.
Right: mass-accretion rate measured at a radius of $500$~km.
All simulations use the common progenitor, SFHo equation of state, and prescribed neutrino-interaction set.}
\label{fig:ccsn_hydro}
\end{figure*}

Figure~\ref{fig:ccsn_hydro} compares the shock radius and the mass-accretion rate at $r=500$~km with the results of the \citet{OConnor2018} comparison.
The additional curves show the six contributing codes: 3DnSNe-IDSA \citep{Takiwaki2014}, AGILE-BOLTZTRAN \citep{Liebendoerfer2004}, FLASH \citep{OConnorCouch2018}, Fornax \citep{Skinner2019}, GR1D \citep{OConnor2015}, and PROMETHEUS-VERTEX \citep{RamppJanka2002}; the green curve denotes NRM1, our existing CPU-based code \citep{Kuroda2016} in one-dimensional Cartoon coordinates, and the red curve denotes \codeName.
We note that, although the computation is initialized from the spherically symmetric progenitor, our full three-dimensional Cartesian code inevitably excites aspherical initial perturbations, which can eventually grow into global-scale convective motions during the post-bounce phase.
Accordingly, the comparison is intended only to assess whether \codeName~can reproduce reasonable post-bounce profiles.
The \codeName~shock radius lies near the upper edge of the inter-code spread, reaching approximately $160$~km at $t_{\rm pb}\simeq70$~ms.
This tendency may plausibly reflect both the three-dimensional nature of the present simulation and its comparatively coarse spatial resolution, which distinguish it from the spherically symmetric reference models and may enhance multidimensional and numerical deviations.

The mass-accretion rate follows the common rapid decline from approximately $7\,M_\odot\,{\rm s}^{-1}$ before bounce to about $2\,M_\odot\,{\rm s}^{-1}$ at the end of the \codeName~time series.
It lies toward the lower side of the comparison band after bounce but reproduces its overall evolution, indicating consistent collapse dynamics and accretion of the progenitor density structure.
At $r=500$~km, where the flow is supersonic and is governed mainly by the gravitational field, the mass-accretion rates obtained with \codeName~and NRM1 differ by approximately $10\%$.
The bounce times also differ between the two simulations by $\sim7$\,ms (and from the remaining results by $\gtrsim10$\,ms), introducing a relative shift along the time axis between their $\dot{M}$ curves and thereby potentially contributing to the apparent discrepancy.
Although the two codes share the same physical formulation and overall numerical structure, \codeName~was independently rewritten from scratch and therefore constitutes a distinct numerical implementation rather than a direct port of NRM1.
Moreover, NRM1 employs a one-dimensional Cartoon method with $N_{\rm cell}=64$ (i.e. $\Delta x=13.4$\,km at $r=500$\,km), whereas the present \codeName~calculation employs a fully three-dimensional Cartesian grid at the lower resolution $N_{\rm cell}=40$ ($\Delta x=21.5$\,km at $r=500$\,km).
Accordingly, a quantitative difference of approximately $10\%$ may be plausible despite the common physical basis of the two codes and should not be interpreted as evidence of inconsistent collapse dynamics.
It should also not be regarded as a GPU-versus-CPU effect; the more relevant differences are the independent implementations, dimensionality, and spatial resolution described above.

\begin{figure*}
\centering
\includegraphics[width=\textwidth]{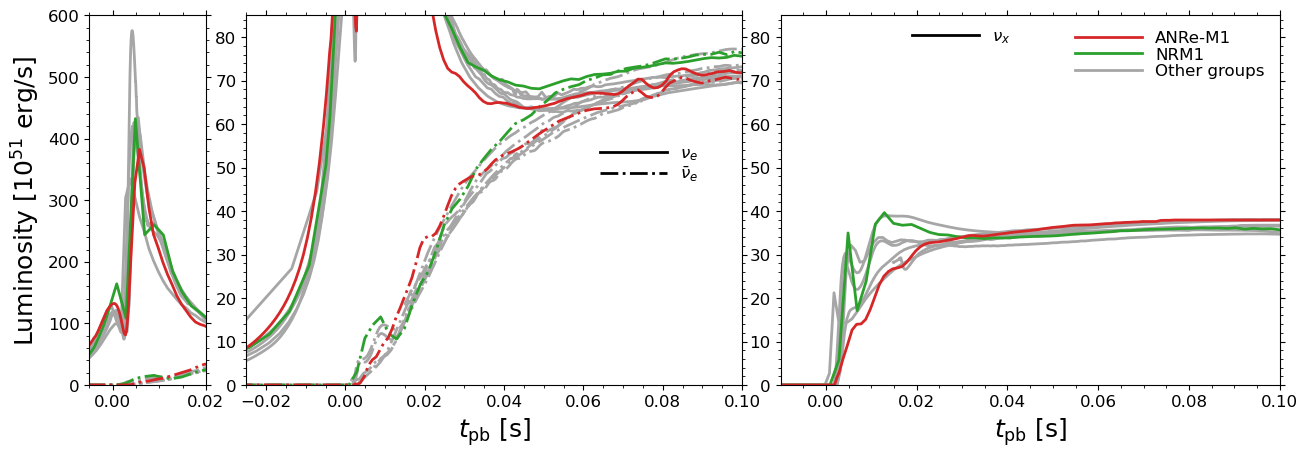}
\caption{Neutrino luminosities as functions of post-bounce time for \codeName (red), the CPU-based NRM1 code (green), and the six other groups participating in the comparison of \citet{OConnor2018}.
The left panel enlarges the prompt electron-neutrino burst, the middle panel shows the electron-neutrino (solid) and electron-antineutrino (dash-dotted) luminosities, and the right panel shows the luminosity of one representative heavy-lepton species.
\label{fig:ccsn_luminosities}}
\end{figure*}
Figure~\ref{fig:ccsn_luminosities} compares the neutrino luminosities with the same reference set.
The electron-neutrino luminosity from \codeName~exhibits the neutronization burst immediately after bounce, with a peak of approximately $3.8\times10^{53}$~erg~s$^{-1}$, followed by the decline toward the accretion-powered phase.
The peak is lower than the largest values among the six other groups but remains within their overall spread. 
After the neutronization burst, the electron-neutrino and electron-antineutrino luminosities approach the comparison band, while the heavy-lepton luminosity rises more gradually and reaches approximately $3.8\times10^{52}$~erg~s$^{-1}$ by $t_{\rm pb}\simeq100$~ms. 
The remaining differences are consistent with the sensitivity of early neutrino signals to the spatial discretization and transport implementation.

\begin{figure*}
\centering
\includegraphics[width=0.9\textwidth]{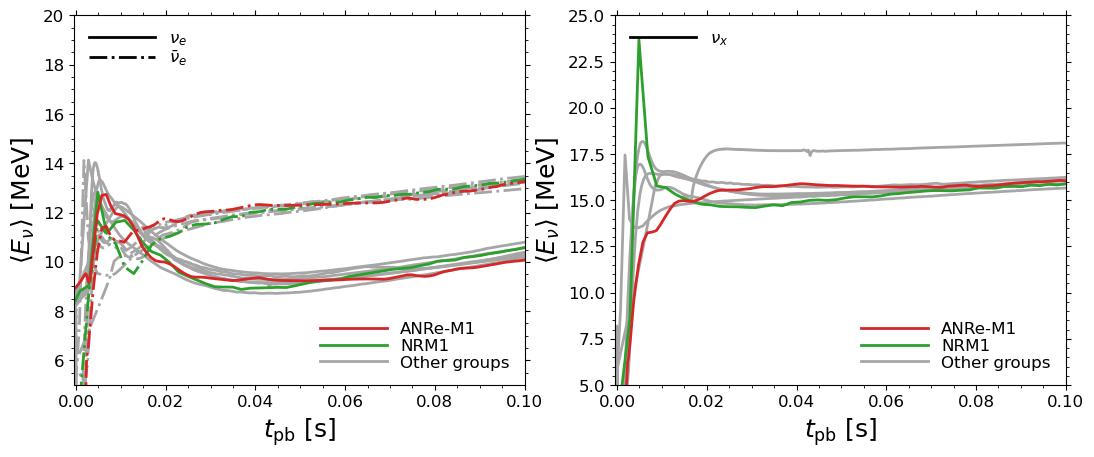}
\caption{Number-weighted mean neutrino energies for \codeName (red), the CPU-based NRM1 code (green), and the six other groups participating in the comparison of \citet{OConnor2018}. The left panel shows electron neutrinos (solid curves) and electron antineutrinos (dash-dotted curves), while the right panel shows one representative heavy-lepton neutrino species.}
\label{fig:ccsn_mean_energies}
\end{figure*}
The corresponding number-weighted mean neutrino energies are shown in Figure~\ref{fig:ccsn_mean_energies}.
The \codeName~electron-flavour mean energies remain within, or close to, the spread of the six other groups after the bounce transient and reproduce the gradual post-bounce rise.
For the heavy-lepton species, \codeName~shows a smoother initial increase than NRM1, which displays a pronounced early overshoot, and reaches the common $15$--$16$~MeV range after approximately $20$~ms. 
Taken together, the hydrodynamic and neutrino comparisons demonstrate that the GPU implementation reproduces the principal features of realistic collapse and early post-bounce evolution while simultaneously evolving the dynamical spacetime, hydrodynamics, and multi-species, multi-energy M1 transport.

\section{Performance}
\label{sec:performance}

\subsection{Systems and benchmark definition}
\label{sec:performance_setup}

We assess the end-to-end performance of \codeName\ using a three-dimensional stellar-collapse calculation that exercises the dynamical spacetime, hydrodynamics, multi-energy M1 transport, and neutrino--matter interaction modules together.
The one-, eight-, and 64-node Viper-GPU calculations use the same progenitor, microphysics, and refinement prescription.
The benchmark employs three neutrino species and \(N_\varepsilon=12\) energy groups, where the latter is a benchmark choice rather than a restriction of the code.
To assess performance for practical CCSN simulations, we employ the more elaborate ``set6a'' opacity set of \citet{Kotake2018}, rather than the baseline set used in Section~\ref{sec:application}.

The GPU measurements were performed on Viper-GPU at the Max Planck Computing and Data Facility (MPCDF).\footnote{\href{https://docs.mpcdf.mpg.de/doc/computing/viper-gpu-user-guide}{MPCDF Viper-GPU}.}
Each Viper-GPU node contains two AMD Instinct MI300A APUs, and each APU combines 24 Zen~4 CPU cores, 228 GPU compute units, and 128~GB of coherent third-generation high-bandwidth memory (HBM3).
The nodes are connected by a non-blocking NDR InfiniBand fat-tree network with a per-node bandwidth of \(400\ {\rm Gb\,s^{-1}}\).
The measurements used eight MPI ranks per node across the two APUs and one host thread per rank.
The executable was built for the Kokkos HIP back-end and the AMD \texttt{gfx942} target with \texttt{-O2} and without an explicit fast-math option, using GCC~14, ROCm~7.0, and Open MPI~5.0 with GPU-aware UCX communication.
\begin{table*}
\centering
\caption{End-to-end weak-scaling performance of \codeName\ on Viper-GPU.
All configurations use \(n_{\rm cell}=20\) and \(N_{\varepsilon}=12\).
The number of computational zones and APUs increases by a factor of eight between successive configurations.
Throughput is defined as the number of zone updates per evolution step divided by \(t_{\rm step}\), and \(E_{\rm weak}\) is normalized to the one-node result.}
\label{tab:performance}
\begin{tabular}{rrrrrrr}
\toprule
\(n_x\) & Nodes & APUs & Zones step\(^{-1}\) & \(t_{\rm step}\) (s) & Zone-cycles s\(^{-1}\) & \(E_{\rm weak}\) (\%)\\
\midrule
2 & 1  & 2   & 704,000    & 0.377186 & \(1.8665\times10^{6}\) & 100.0\\
4 & 8  & 16  & 5,632,000  & 0.478010 & \(1.1782\times10^{7}\) & 78.9\\
8 & 64 & 128 & 45,056,000 & 0.608274 & \(7.4072\times10^{7}\) & 62.0\\
\bottomrule
\end{tabular}
\end{table*}

\begin{figure}
\centering
\includegraphics[width=\columnwidth]{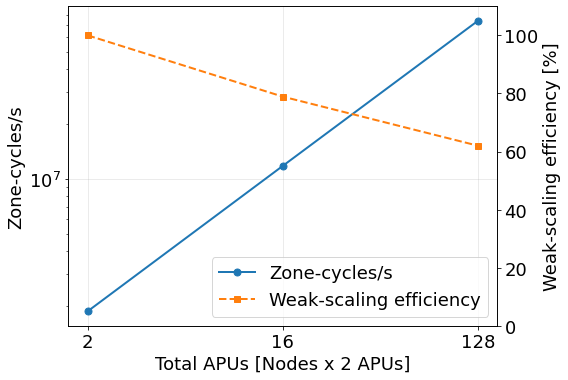}
\caption{Weak-scaling performance of \codeName\ on Viper-GPU.
The blue curve shows the aggregate throughput in zone-cycles per second, while the orange curve shows the weak-scaling efficiency normalized to the two-APU result.
The workload per APU is held fixed as the allocation increases from 2 to 128 APUs.}
\label{fig:weak_scaling}
\end{figure}

\subsection{Weak scaling}
\label{sec:weak_scaling}

The weak-scaling test increases \(n_x=n_y=n_z\) from 2 to 4 and 8 and simultaneously increases the Viper-GPU allocation from 1 to 8 and 64 nodes, respectively, assuming an octant symmetry.
The corresponding APU counts are 2, 16, and 128.
Each MPI rank owns one subblock containing $n_{\rm cell}^3=20^3$ active cells at every FMR level, with the levels running from $l=0$ to $l=L_{\rm FMR}-1=10$.
The number of subblocks, computational zones, and APUs therefore grows by a factor of eight at each step, leaving the workload per APU fixed.
Because the number of computational zones (i.e. $n_x\times n_y\times n_z\times n_{\rm cell}^3\times L_{\rm FMR}$) increases from 704,000 to 5,632,000 and 45,056,000, the workload per APU remains constant at 352,000 zones per evolution step.
We define the weak-scaling efficiency by
\begin{equation}
 E_{\rm weak}(P)
 =\frac{t_{\rm step}(P_0)}{t_{\rm step}(P)},
\label{eq:weak_scaling_efficiency}
\end{equation}
where \(P_0=2\) APUs on one Viper-GPU node.

As shown in Table~\ref{tab:performance} and Figure~\ref{fig:weak_scaling}, the step time increases from 0.377186 to 0.478010 and 0.608274~s as the allocation grows from 1 to 8 and 64 nodes.
The corresponding weak-scaling efficiencies are 78.9 and 62.0 per cent.
At the largest scale, the calculation advances 45.056 million zones per step and sustains \(7.4072\times10^7\) zone-cycles~s\(^{-1}\), which is 39.7 times the aggregate throughput of the one-node calculation while using 64 times as many APUs.

The aggregate rates in Table~\ref{tab:performance} correspond to \(9.33\times10^5\), \(7.36\times10^5\), and \(5.79\times10^5\) zone-cycles~s\(^{-1}\)~APU\(^{-1}\) at 2, 16, and 128 APUs, respectively.
These values are lower than those reported for GPU-accelerated numerical-relativity and GRMHD codes with less expensive per-zone physics.
For example, the dynamical-spacetime production setup of GRaM-X reaches \(4.12\times10^6\) zone-cycles~s\(^{-1}\)~GPU\(^{-1}\) on six NVIDIA V100 GPUs, while AthenaK reports \(1.07\times10^7\) zone-cycles~s\(^{-1}\) on a single NVIDIA A100 GPU for a freely evolving relativistic-star test \citep{Shankar2023,Zhu2025}.
Thus, depending on the code, hardware, and benchmark, the per-accelerator throughput of \codeName\ is lower by a factor of several to approximately one order of magnitude.
This comparison is indicative rather than controlled because the algorithms, accelerator architectures, refinement structures, and definitions of a complete evolution step differ.

The lower raw zone rate is consistent with the substantially larger amount of physics performed in each \codeName\ zone-cycle.
With three neutrino species and \(N_\varepsilon=12\) energy groups, spatial transport advances \(N_\nu N_\varepsilon\times4=144\) radiation-moment components per cell in addition to the hydrodynamic and spacetime variables.
The stiff neutrino--matter coupling is then updated by a Newton iteration whose local unknown vector contains these $N_\nu N_\varepsilon\times4(=144)$ radiation variables and five hydrodynamic primitive variables.
The implementation therefore constructs and solves a Jacobian system of dimension $(N_\nu N_\varepsilon\times4+5)\times(N_\nu N_\varepsilon\times4+5)=149\times149$ in every active cell during the implicit source update.
A zone-cycle metric assigns unit weight to this complete operation and does not normalize for the number of evolved moments or the local nonlinear solve.
Consequently, the factor-of-several to order-of-magnitude difference from pure numerical-relativity or GRMHD benchmarks is compatible with the intended production supernova workload and should not by itself be interpreted as poor accelerator utilization.

The loss of weak-scaling efficiency, by contrast, remains an important optimization target.
A likely leading cause is the small active block assigned to each MPI rank, \(n_{\rm cell}^3=20^3\), which is constrained by the large memory footprint of the multi-energy radiation variables and implicit-coupling workspace (i.e. the Jacobian matrix).
For a cubic block, the face-exchange volume scales as \(n_{\rm cell}^2\), whereas the active volume scales as \(n_{\rm cell}^3\), so their ratio is proportional to \(n_{\rm cell}^{-1}\).
The communication-to-computation ratio is consequently much larger than for the \(64^3\)--\(240^3\) blocks or per-accelerator domains commonly used in other scaling studies \citep{Shankar2023,Zhu2025}.
As the node count grows, a larger fraction of these ghost-zone exchanges crosses the inter-node network, so communication, synchronization, and FMR-boundary operations consume an increasing fraction of each step.
Similar workload dependence was reported by GRaM-X, whose efficiency decreased when the cells per GPU were reduced and whose AMR case incurred additional prolongation, restriction, and coarse--fine communication costs \citep{Shankar2023}.

Increasing the active block size or assigning fewer, larger subdomains to each APU should improve the surface-to-volume ratio, provided that the radiation and Newton-solver memory footprint can be reduced sufficiently to avoid exhausting HBM.
Complementary priorities are overlapping ghost-zone exchange with independent kernels, aggregating small messages, optimizing rank-to-APU placement, and profiling communication and refinement-level load balance.
The present cumulative timers do not isolate these components, so dedicated MPI and kernel profiles are required to quantify their individual contributions.

\subsection{Comparison with the original Fortran code}
\label{sec:fortran_performance}

For a practical comparison with the established CPU workflow, the same setup described in Section~\ref{sec:performance_setup} was evolved using our existing Fortran code (NRM1) on one node of the SAKURA system introduced in Section~\ref{sec:TOV star}.
The Fortran calculation used the same nominal resolution parameters as the one-node Viper-GPU calculation, namely \(n_x=2\), \(n_{\rm cell}=20\), and \(N_\varepsilon=12\).
The run used a hybrid MPI--OpenMP configuration with eight MPI ranks on the node and four OpenMP threads per rank, corresponding to 32 CPU cores in total.
We measure the step time as
\begin{equation}
 t_{\rm step}^{\rm NRM1}
 =24.98\ {\rm s\ step^{-1}}.
\label{eq:fortran_tstep}
\end{equation}
Comparing this value with \(t_{\rm step}=0.377186\ {\rm s\ step^{-1}}\) for the one-node Viper-GPU benchmark (see Table~\ref{tab:performance}) gives
\begin{equation}
 {\cal S}_{\rm NRM1\rightarrow \codeName}
 =\frac{t_{\rm step}^{\rm NRM1}}
 {t_{\rm step}^{\codeName}}
 \simeq66.2.
\label{eq:fortran_gpu_speedup}
\end{equation}
The GPU implementation therefore provides an indicative reduction of an order of magnitude in the wall-clock time per step relative to the established Fortran workflow.

This factor should, however, be regarded as an order-of-magnitude indication rather than a controlled code-to-code benchmark.
The two runs differ in processor architecture, memory system, compiler toolchain, and programming model, with the original code written in Fortran and the accelerated implementation written in C++ using Kokkos.
Consequently, the factor of approximately 66 reflects the combined effects of these hardware and software differences and cannot be interpreted at face value as an isolated speed-up due to either Kokkos or the MI300A architecture.
A controlled performance assessment would require otherwise equivalent implementations and matched numerical workloads on the systems being compared.

\section{Conclusions}\label{sec:conclusions}

We have presented \codeName, a performance-portable numerical-relativity code for multidimensional core-collapse supernova simulations with multi-energy M1 neutrino transport.
The code retains the general-relativistic radiation-hydrodynamics formulation and neutrino--matter interactions of our established Fortran framework while reorganizing its data structures, kernels, and parallel decomposition for accelerator execution through Kokkos.
The principal evolution data remain resident in device memory, and the computationally intensive hydrodynamic, spacetime, neutrino-transport, and implicit source-term kernels are expressed using portable parallel patterns.

We verified the new implementation with a suite of problems spanning relativistic hydrodynamics, dynamical spacetime evolution, and neutrino transport.
The shock-tube, linear-wave, equilibrium-star, adiabatic stellar-collapse, and radiation-transport tests reproduce the expected solutions and convergence behaviour.
These tests collectively exercise the principal components required for self-consistent core-collapse simulations.

As an integrated application, we followed stellar collapse, core bounce, and the early post-bounce evolution using energy-dependent neutrino transport.
To assess the reliability of \codeName, particularly its neutrino-transport implementation, we adopted the same simulation setup as the comparison study of \citet{OConnor2018} and performed detailed comparisons of the resulting physical quantities.
During collapse, \codeName~reproduces the expected onset of neutrino trapping when the central density reaches $\rho_c\sim10^{12}$\,g\,cm$^{-3}$.
The simulation subsequently reaches core bounce at approximately 291 ms after the onset of collapse, consistent with the range reported in the multi-code comparison of \citet{OConnor2018}.
During the early post-bounce phase, the evolution of the neutrino luminosities, mean energies, and shock radius is broadly consistent with the one-dimensional results reported by \citet{OConnor2018} and with our original code NRM1.
This agreement demonstrates that the coupled transport, weak interactions, hydrodynamics, and relativistic gravity operate coherently in the accelerator implementation, although longer and higher-resolution multidimensional simulations will be required to assess the fully developed post-bounce dynamics.

On the AMD MI300A system, the representative production-like workload reaches approximately \(10^6\) zone-cycles~s\(^{-1}\) per APU.
The weak-scaling efficiency is 78.9 per cent on 16 APUs and 62.0 per cent on 128 APUs relative to the two-APU case.
The measured step time is approximately 66 times shorter than for the available run of the original Fortran code on the CPU cluster, but this value combines differences in processor architecture, memory system, compiler toolchain, and programming model and should therefore be regarded as an indicative practical comparison rather than a controlled acceleration factor.
The decline in weak-scaling efficiency identifies inter-rank communication and the small active volume per MPI rank as important optimization targets.

Future development will reduce the memory footprint of the multi-energy radiation fields and implicit solver so that larger active blocks can be assigned to each accelerator.
Further priorities include overlapping communication with independent kernels, reducing synchronization and small-message overheads, improving refinement-level load balance, and extending performance tests to additional accelerator architectures.
Together with longer-duration production simulations, these developments will establish \codeName\ as a portable framework for multidimensional general-relativistic core-collapse supernova modelling on present and forthcoming heterogeneous supercomputers.

\section*{Acknowledgements}
TK is grateful to Ming-Zhe Han and Carlo Musolino for fruitful discussions on GPU implementation.
Numerical computations were carried out on SAKURA and Viper-GPU at the Max Planck Computing and Data Facility (MPCDF).
This work was supported in part by JSPS KAKENHI Grant Number 23H04900.

\section*{Data availability}
The data underlying this article will be shared on reasonable request to the corresponding author.
The availability of the source code and analysis scripts will be specified in the accepted version of the manuscript.

\bibliographystyle{mnras}
\bibliography{references}

@article{Edwards2014,
 author={Edwards, H. C. and Trott, C. R. and Sunderland, D.},
 title={Kokkos: Enabling manycore performance portability through polymorphic memory access patterns},
 journal={Journal of Parallel and Distributed Computing}, year={2014}, volume={74}, pages={3202--3216},
 doi={10.1016/j.jpdc.2014.07.003}}

@ARTICLE{Janka2025_review,
       author = {{Janka}, Hans-Thomas},
        title = "{Long-Term Multidimensional Models of Core-Collapse Supernovae: Progress and Challenges}",
      journal = {Annual Review of Nuclear and Particle Science},
         year = 2025,
        month = sep,
       volume = {75},
       number = {1},
        pages = {425-461},
          doi = {10.1146/annurev-nucl-121423-100945},
archivePrefix = {arXiv},
       eprint = {2502.14836},
 primaryClass = {astro-ph.HE},
       adsurl = {https://ui.adsabs.harvard.edu/abs/2025ARNPS..75..425J}
}

@ARTICLE{Yamada2024,
       author = {{Yamada}, Shoichi and {Nagakura}, Hiroki and {Akaho}, Ryuichiro and {Harada}, Akira and {Furusawa}, Shun and {Iwakami}, Wakana and {Okawa}, Hirotada and {Matsufuru}, Hideo and {Sumiyoshi}, Kohsuke},
        title = "{Physical mechanism of core-collapse supernovae that neutrinos drive}",
      journal = {Proceedings of the Japan Academy, Series B},
         year = 2024,
        month = mar,
       volume = {100},
       number = {3},
        pages = {190-233},
          doi = {10.2183/pjab.100.015},
       adsurl = {https://ui.adsabs.harvard.edu/abs/2024PJAB..100..190Y}
}

@ARTICLE{Fryer2023,
       author = {{Fryer}, Christopher L. and {Burns}, Eric and {Hungerford}, Aimee and {Safi-Harb}, Samar and {Wollaeger}, R.~T. and {Miller}, Richard S. and {Negro}, Michela and {Anandagoda}, Samalka and {Hartmann}, Dieter H.},
        title = "{Multimessenger Diagnostics of the Engine behind Core-collapse Supernovae}",
      journal = {\apj},
         year = 2023,
        month = oct,
       volume = {956},
       number = {1},
          eid = {19},
        pages = {19},
          doi = {10.3847/1538-4357/ace0c3},
archivePrefix = {arXiv},
       eprint = {2305.06134},
 primaryClass = {astro-ph.HE},
       adsurl = {https://ui.adsabs.harvard.edu/abs/2023ApJ...956...19F}
}

@INPROCEEDINGS{Mezzacappa2023,
       author = {{Mezzacappa}, Anthony},
        title = "{Toward Realistic Models of Core Collapse Supernovae: A Brief Review}",
    booktitle = {The Predictive Power of Computational Astrophysics as a Discover Tool},
         year = 2023,
       editor = {{Bisikalo}, Dmitry and {Wiebe}, Dmitri and {Boily}, Christian},
       series = {IAU Symposium},
       volume = {362},
        month = jan,
        pages = {215-227},
          doi = {10.1017/S1743921322001831},
archivePrefix = {arXiv},
       eprint = {2205.13438},
 primaryClass = {astro-ph.SR},
       adsurl = {https://ui.adsabs.harvard.edu/abs/2023IAUS..362..215M}
}

@ARTICLE{Burrows2021_review,
       author = {{Burrows}, A. and {Vartanyan}, D.},
        title = "{Core-collapse supernova explosion theory}",
      journal = {\nat},
         year = 2021,
        month = jan,
       volume = {589},
       number = {7840},
        pages = {29-39},
          doi = {10.1038/s41586-020-03059-w},
archivePrefix = {arXiv},
       eprint = {2009.14157},
 primaryClass = {astro-ph.SR},
       adsurl = {https://ui.adsabs.harvard.edu/abs/2021Natur.589...29B}
}

@article{Kuroda2016,
 author={Kuroda, T. and Takiwaki, T. and Kotake, K.},
 title={A new multi-energy neutrino radiation-hydrodynamics code in full general relativity and its application to the gravitational collapse of massive stars},
 journal={The Astrophysical Journal Supplement Series}, year={2016}, volume={222}, number={2}, pages={20},
 doi={10.3847/0067-0049/222/2/20}}

@article{Bernuzzi2010,
 author={Bernuzzi, S. and Hilditch, D.},
 title={Constraint violation in free evolution schemes: Comparing the {BSSNOK} formulation with a conformal decomposition of the {Z4} formulation},
 journal={Physical Review D}, year={2010}, volume={81}, number={8}, pages={084003},
 doi={10.1103/PhysRevD.81.084003}}

@article{Hilditch2013,
 author={Hilditch, D. and Bernuzzi, S. and Thierfelder, M. and Cao, Z. and Tichy, W. and Br{\"u}gmann, B.},
 title={Compact binary evolutions with the {Z4c} formulation},
 journal={Physical Review D}, year={2013}, volume={88}, number={8}, pages={084057},
 doi={10.1103/PhysRevD.88.084057}}

@article{Thorne1981,
 author={Thorne, K. S.},
 title={Relativistic radiative transfer: moment formalisms},
 journal={Monthly Notices of the Royal Astronomical Society}, year={1981}, volume={194}, number={2}, pages={439--473},
 doi={10.1093/mnras/194.2.439}}

@article{Shibata2011,
 author={Shibata, M. and Kiuchi, K. and Sekiguchi, Y. and Suwa, Y.},
 title={General relativistic radiation magnetohydrodynamics: formulation and code tests},
 journal={Progress of Theoretical Physics}, year={2011}, volume={125}, number={6}, pages={1255--1287},
 doi={10.1143/PTP.125.1255}}

@article{Minerbo1978,
 author={Minerbo, G. N.},
 title={Maximum entropy Eddington factors},
 journal={Journal of Quantitative Spectroscopy and Radiative Transfer}, year={1978}, volume={20}, number={6}, pages={541--545},
 doi={10.1016/0022-4073(78)90024-9}}

@article{Trott2022,
 author={Trott, C. R. and others}, title={Kokkos 3: Programming model extensions for the exascale era},
 journal={IEEE Transactions on Parallel and Distributed Systems}, year={2022}, volume={33}, number={4}, pages={805--817},
 doi={10.1109/TPDS.2021.3097283}}

@article{Mueller2012,
 author={M{\"u}ller, B. and Janka, H.-T. and Marek, A.},
 title={A new multi-dimensional general relativistic neutrino hydrodynamics code for core-collapse supernovae. II. Relativistic explosion models of core-collapse supernovae},
 journal={The Astrophysical Journal}, year={2012}, volume={756}, number={1}, pages={84},
 doi={10.1088/0004-637X/756/1/84}}

@article{Mueller2017,
 author={M{\"u}ller, B. and Melson, T. and Heger, A. and Janka, H.-T.},
 title={Supernova simulations from a 3D progenitor model: impact of perturbations and evolution of explosion properties},
 journal={Monthly Notices of the Royal Astronomical Society}, year={2017}, volume={472}, number={1}, pages={491--513},
 doi={10.1093/mnras/stx1962}}

@article{Skinner2019,
 author={Skinner, M. A. and Dolence, J. C. and Burrows, A. and Radice, D. and Vartanyan, D.},
 title={{Fornax}: A flexible code for multiphysics astrophysical simulations},
 journal={The Astrophysical Journal Supplement Series}, year={2019}, volume={241}, number={1}, pages={7},
 doi={10.3847/1538-4365/ab007f}}

@article{Burrows2020,
 author={Burrows, A. and Radice, D. and Vartanyan, D. and Nagakura, H. and Skinner, M. A. and Dolence, J. C.},
 title={The overarching framework of core-collapse supernova explosions as revealed by 3D {Fornax} simulations},
 journal={Monthly Notices of the Royal Astronomical Society}, year={2020}, volume={491}, number={2}, pages={2715--2735},
 doi={10.1093/mnras/stz3223}}

@article{Just2015,
 author={Just, O. and Obergaulinger, M. and Janka, H.-T.},
 title={A new multidimensional, energy-dependent two-moment transport code for neutrino-hydrodynamics},
 journal={Monthly Notices of the Royal Astronomical Society}, year={2015}, volume={453}, number={4}, pages={3386--3413},
 doi={10.1093/mnras/stv1892}}

@article{Bruenn2020,
 author={Bruenn, S. W. and Blondin, J. M. and Hix, W. R. and Lentz, E. J. and Messer, O. E. B. and Mezzacappa, A. and Endeve, E. and Harris, J. A. and Marronetti, P. and Budiardja, R. D. and Chertkow, M. A. and Lee, C.-T.},
 title={{CHIMERA}: A massively parallel code for core-collapse supernova simulations},
 journal={The Astrophysical Journal Supplement Series}, year={2020}, volume={248}, number={1}, pages={11},
 doi={10.3847/1538-4365/ab7aff}}

@article{Nagakura2018,
 author={Nagakura, H. and Iwakami, W. and Furusawa, S. and Okawa, H. and Harada, A. and Sumiyoshi, K. and Yamada, S. and Matsufuru, H. and Imakura, A.},
 title={Simulations of core-collapse supernovae in spatial axisymmetry with full Boltzmann neutrino transport},
 journal={The Astrophysical Journal}, year={2018}, volume={854}, number={2}, pages={136},
 doi={10.3847/1538-4357/aaac29}}

@article{Akaho2021,
 author={Akaho, R. and Harada, A. and Nagakura, H. and Sumiyoshi, K. and Iwakami, W. and Okawa, H. and Furusawa, S. and Matsufuru, H. and Yamada, S.},
 title={Multidimensional Boltzmann neutrino transport code in full general relativity for core-collapse simulations},
 journal={The Astrophysical Journal}, year={2021}, volume={909}, number={2}, pages={210},
 doi={10.3847/1538-4357/abe1bf}}

@article{Bollig2021,
 author={Bollig, R. and Yadav, N. and Kresse, D. and Janka, H.-T. and M{\"u}ller, B. and Heger, A.},
 title={Self-consistent 3D supernova models from -7 minutes to +7 s: a 1-bethe explosion of a \({\sim}19\,{\rm M}_{\odot}\) progenitor},
 journal={The Astrophysical Journal}, year={2021}, volume={915}, number={1}, pages={28},
 doi={10.3847/1538-4357/abf82e}}

@article{Wang2023,
 author={Wang, T. and Burrows, A.},
 title={Neutrino-driven winds in three-dimensional core-collapse supernova simulations},
 journal={The Astrophysical Journal}, year={2023}, volume={954}, number={2}, pages={114},
 doi={10.3847/1538-4357/ace7b2}}

@article{Barker2024,
 author={Barker, B. and Gogilashvili, M. and Rodriguez-Bueno, J. and Fields, C. and Dolence, J. and Miller, J. and Murphy, J. and Roberts, L. and Ryan, B.},
 title={{Phoebus}: Performance portable {GRRMHD} for relativistic astrophysics},
 journal={arXiv e-prints}, year={2024}, pages={arXiv:2410.09146},
 doi={10.48550/arXiv.2410.09146},
 archivePrefix={arXiv}, eprint={2410.09146}}

@article{Shankar2023,
 author={Shankar, S. and M{\"o}sta, P. and Brandt, S. R. and Haas, R. and Schnetter, E. and de Graaf, Y.},
 title={{GRaM-X}: A new GPU-accelerated dynamical spacetime {GRMHD} code for exascale computing with the {Einstein Toolkit}},
 journal={Classical and Quantum Gravity}, year={2023}, volume={40}, number={20}, pages={205009},
 doi={10.1088/1361-6382/acf2d9}}

@article{Zhu2025,
 author={Zhu, H. and Fields, J. and Zappa, F. and Radice, D. and Stone, J. M. and Rashti, A. and Cook, W. and Bernuzzi, S. and Daszuta, B.},
 title={Performance-portable numerical relativity with {AthenaK}},
 journal={The Astrophysical Journal Supplement Series}, year={2025}, volume={276}, number={2}, pages={50},
 doi={10.3847/1538-4365/adcf96}}

@article{Kalinani2025,
 author={Kalinani, J. V. and Ji, L. and Ennoggi, L. and Lopez Armengol, F. G. and Timotheo Sanches, L. and Tsao, B.-J. and Brandt, S. R. and Campanelli, M. and Ciolfi, R. and Giacomazzo, B. and Haas, R. and Schnetter, E. and Zlochower, Y.},
 title={{AsterX}: A new open-source GPU-accelerated {GRMHD} code for dynamical spacetimes},
 journal={Classical and Quantum Gravity}, year={2025}, volume={42}, number={2}, pages={025016},
 doi={10.1088/1361-6382/ad9c11}}

@article{Han2026,
 author={Han, M.-Z. and Kiuchi, K. and Shibata, M.},
 title={{SACRA-K}: A performance-portable numerical relativity code with {Kokkos}},
 journal={arXiv e-prints}, year={2026}, pages={arXiv:2607.08743},
 doi={10.48550/arXiv.2607.08743},
 archivePrefix={arXiv}, eprint={2607.08743}}

@article{Musolino2026,
 author={Musolino, C. and Ecker, C. and Topolski, K. and Cassing, M. and Miler, K. and Ng, H. H.-Y. and Pierre, K. and Most, E. R. and Rezzolla, L.},
 title={{GRACE}: An open-source framework for GPU-accelerated numerical relativity},
 journal={arXiv e-prints}, year={2026}, pages={arXiv:2607.09854},
 doi={10.48550/arXiv.2607.09854},
 archivePrefix={arXiv}, eprint={2607.09854}}

@article{Pons2000,
    author = {Pons, J. A. and Ibáñez, J. Ma. and Miralles, J. A.},
    title = {Hyperbolic character of the angular moment equations of radiative transfer and numerical methods},
    journal = {Monthly Notices of the Royal Astronomical Society},
    volume = {317},
    number = {3},
    pages = {550-562},
    year = {2000},
    month = {09},
    issn = {0035-8711},
    doi = {10.1046/j.1365-8711.2000.03679.x},
    url = {https://doi.org/10.1046/j.1365-8711.2000.03679.x},
    eprint = {https://academic.oup.com/mnras/article-pdf/317/3/550/2894911/317-3-550.pdf},
}

@ARTICLE{Audit2002,
       author = {{Audit}, E. and {Charrier}, P. and {Chi{\`e}ze}, J.  -P. and {Dubroca}, B.},
        title = "{A radiation-hydrodynamics scheme valid from the transport to the diffusion limit}",
      journal = {arXiv e-prints},
         year = 2002,
        month = jun,
          eid = {astro-ph/0206281},
        pages = {astro-ph/0206281},
          doi = {10.48550/arXiv.astro-ph/0206281},
archivePrefix = {arXiv},
       eprint = {astro-ph/0206281},
 primaryClass = {astro-ph},
       adsurl = {https://ui.adsabs.harvard.edu/abs/2002astro.ph..6281A}
}

@ARTICLE{OConnorOtt2013,
       author = {{O'Connor}, Evan and {Ott}, Christian D.},
        title = "{The Progenitor Dependence of the Pre-explosion Neutrino Emission in Core-collapse Supernovae}",
      journal = {\apj},
         year = 2013,
        month = jan,
       volume = {762},
       number = {2},
          eid = {126},
        pages = {126},
          doi = {10.1088/0004-637X/762/2/126},
archivePrefix = {arXiv},
       eprint = {1207.1100},
 primaryClass = {astro-ph.HE},
       adsurl = {https://ui.adsabs.harvard.edu/abs/2013ApJ...762..126O}
}

@ARTICLE{Mignone2005,
       author = {{Mignone}, A. and {Bodo}, G.},
        title = "{An HLLC Riemann solver for relativistic flows - I. Hydrodynamics}",
      journal = {\mnras},
         year = 2005,
        month = nov,
       volume = {364},
       number = {1},
        pages = {126-136},
          doi = {10.1111/j.1365-2966.2005.09546.x},
archivePrefix = {arXiv},
       eprint = {astro-ph/0506414},
 primaryClass = {astro-ph},
       adsurl = {https://ui.adsabs.harvard.edu/abs/2005MNRAS.364..126M}
}

@ARTICLE{OConnorOtt2011,
       author = {{O'Connor}, Evan and {Ott}, Christian D.},
        title = "{Black Hole Formation in Failing Core-Collapse Supernovae}",
      journal = {\apj},
         year = 2011,
        month = apr,
       volume = {730},
       number = {2},
          eid = {70},
        pages = {70},
          doi = {10.1088/0004-637X/730/2/70},
archivePrefix = {arXiv},
       eprint = {1010.5550},
 primaryClass = {astro-ph.HE},
       adsurl = {https://ui.adsabs.harvard.edu/abs/2011ApJ...730...70O}
}

@ARTICLE{Kotake2018,
       author = {{Kotake}, Kei and {Takiwaki}, Tomoya and {Fischer}, Tobias and
         {Nakamura}, Ko and {Mart{\'\i}nez-Pinedo}, Gabriel},
        title = "{Impact of Neutrino Opacities on Core-collapse Supernova Simulations}",
      journal = {\apj},
         year = "2018",
        month = "Feb",
       volume = {853},
       number = {2},
          eid = {170},
        pages = {170},
          doi = {10.3847/1538-4357/aaa716},
archivePrefix = {arXiv},
       eprint = {1801.02703},
 primaryClass = {astro-ph.HE},
       adsurl = {https://ui.adsabs.harvard.edu/abs/2018ApJ...853..170K}
}

@ARTICLE{UmedaNomoto2008,
       author = {{Umeda}, Hideyuki and {Nomoto}, Ken'ichi},
        title = "{How Much $^{56}$Ni Can Be Produced in Core-Collapse Supernovae? Evolution and Explosions of 30-100 M$_{☉}$ Stars}",
      journal = {\apj},
         year = 2008,
        month = feb,
       volume = {673},
       number = {2},
        pages = {1014-1022},
          doi = {10.1086/524767},
archivePrefix = {arXiv},
       eprint = {0707.2598},
 primaryClass = {astro-ph},
       adsurl = {https://ui.adsabs.harvard.edu/abs/2008ApJ...673.1014U}
}

@ARTICLE{MartiMueller2003,
       author = {{Mart{\'\i}}, Jos{\'e} Maria and {M{\"u}ller}, Ewald},
        title = "{Numerical Hydrodynamics in Special Relativity}",
      journal = {Living Reviews in Relativity},
         year = 2003,
        month = dec,
       volume = {6},
       number = {1},
          eid = {7},
        pages = {7},
          doi = {10.12942/lrr-2003-7},
       adsurl = {https://ui.adsabs.harvard.edu/abs/2003LRR.....6....7M}
}

@ARTICLE{SFH,
   author = {{Steiner}, A.~W. and {Hempel}, M. and {Fischer}, T.},
    title = "{Core-collapse Supernova Equations of State Based on Neutron Star Observations}",
  journal = {\apj},
archivePrefix = "arXiv",
   eprint = {1207.2184},
 primaryClass = "astro-ph.SR",
     year = 2013,
    month = sep,
   volume = 774,
      eid = {17},
    pages = {17},
      doi = {10.1088/0004-637X/774/1/17},
   adsurl = {http://adsabs.harvard.edu/abs/2013ApJ...774...17S}
}

@ARTICLE{Couch2015,
  author = {{Couch}, Sean M. and {Ott}, Christian D.},
  title = "{The Role of Turbulence in Neutrino-driven Core-collapse Supernova Explosions}",
  journal = {The Astrophysical Journal},
  year = {2015},
  month = {jan},
  volume = {799},
  number = {1},
  eid = {5},
  pages = {5},
  doi = {10.1088/0004-637X/799/1/5},
  eprint = {1408.1399},
  archivePrefix = {arXiv},
  primaryClass = {astro-ph.HE},
  adsurl = {https://ui.adsabs.harvard.edu/abs/2015ApJ...799....5C}
}

@ARTICLE{Summa2018,
  author = {{Summa}, Alexander and {Janka}, Hans-Thomas and {Melson}, Tobias and {Marek}, Andreas},
  title = "{Rotation-supported Neutrino-driven Supernova Explosions in Three Dimensions and the Critical Luminosity Condition}",
  journal = {The Astrophysical Journal},
  year = {2018},
  month = {jan},
  volume = {852},
  number = {1},
  eid = {28},
  pages = {28},
  doi = {10.3847/1538-4357/aa9ce8},
  eprint = {1708.04154},
  archivePrefix = {arXiv},
  primaryClass = {astro-ph.HE},
  adsurl = {https://ui.adsabs.harvard.edu/abs/2018ApJ...852...28S}
}

@ARTICLE{WH2007,
       author = {{Woosley}, S.~E. and {Heger}, A.},
        title = "{Nucleosynthesis and remnants in massive stars of solar metallicity}",
      journal = {\physrep},
         year = 2007,
        month = apr,
       volume = {442},
       number = {1-6},
        pages = {269-283},
          doi = {10.1016/j.physrep.2007.02.009},
archivePrefix = {arXiv},
       eprint = {astro-ph/0702176},
 primaryClass = {astro-ph},
       adsurl = {https://ui.adsabs.harvard.edu/abs/2007PhR...442..269W}
}

@ARTICLE{Fujibayashi2020,
  author = {{Fujibayashi}, Sho and {Wanajo}, Shinya and {Kiuchi}, Kenta and {Kyutoku}, Koutarou and {Sekiguchi}, Yuichiro and {Shibata}, Masaru},
  title = "{Postmerger Mass Ejection of Low-mass Binary Neutron Stars}",
  journal = {The Astrophysical Journal},
  year = {2020},
  month = {oct},
  volume = {901},
  number = {2},
  eid = {122},
  pages = {122},
  doi = {10.3847/1538-4357/abafc2},
  eprint = {2007.00474},
  archivePrefix = {arXiv},
  primaryClass = {astro-ph.HE},
  adsurl = {https://ui.adsabs.harvard.edu/abs/2020ApJ...901..122F}
}

@ARTICLE{Radice2016,
  author = {{Radice}, David and {Ott}, Christian D. and {Abdikamalov}, Ernazar and {Couch}, Sean M. and {Haas}, Roland and {Schnetter}, Erik},
  title = "{Neutrino-driven Convection in Core-collapse Supernovae: High-resolution Simulations}",
  journal = {The Astrophysical Journal},
  year = {2016},
  month = {mar},
  volume = {820},
  number = {1},
  eid = {76},
  pages = {76},
  doi = {10.3847/0004-637X/820/1/76},
  eprint = {1510.05022},
  archivePrefix = {arXiv},
  primaryClass = {astro-ph.HE},
  adsurl = {https://ui.adsabs.harvard.edu/abs/2016ApJ...820...76R}
}

@ARTICLE{Nagakura2019,
  author = {{Nagakura}, Hiroki and {Burrows}, Adam and {Radice}, David and {Vartanyan}, David},
  title = "{Towards an understanding of the resolution dependence of core-collapse supernova simulations}",
  journal = {Monthly Notices of the Royal Astronomical Society},
  year = {2019},
  month = {dec},
  volume = {490},
  number = {4},
  pages = {4622--4637},
  doi = {10.1093/mnras/stz2730},
  eprint = {1905.03786},
  archivePrefix = {arXiv},
  primaryClass = {astro-ph.HE},
  adsurl = {https://ui.adsabs.harvard.edu/abs/2019MNRAS.490.4622N}
}

@ARTICLE{Walk2018,
  author = {{Walk}, Laurie and {Tamborra}, Irene and {Janka}, Hans-Thomas and {Summa}, Alexander},
  title = "{Identifying rotation in SASI-dominated core-collapse supernovae with a neutrino gyroscope}",
  journal = {Physical Review D},
  year = {2018},
  month = {dec},
  volume = {98},
  number = {12},
  eid = {123001},
  pages = {123001},
  doi = {10.1103/PhysRevD.98.123001},
  eprint = {1807.02366},
  archivePrefix = {arXiv},
  primaryClass = {astro-ph.HE},
  adsurl = {https://ui.adsabs.harvard.edu/abs/2018PhRvD..98l3001W}
}

@ARTICLE{Radice2018,
  author = {{Radice}, David and {Abdikamalov}, Ernazar and {Ott}, Christian D. and {M{\"o}sta}, Philipp and {Couch}, Sean M. and {Roberts}, Luke F.},
  title = "{Turbulence in core-collapse supernovae}",
  journal = {Journal of Physics G: Nuclear and Particle Physics},
  year = {2018},
  month = {may},
  volume = {45},
  number = {5},
  eid = {053003},
  pages = {053003},
  doi = {10.1088/1361-6471/aab872},
  eprint = {1710.01282},
  archivePrefix = {arXiv},
  primaryClass = {astro-ph.HE},
  adsurl = {https://ui.adsabs.harvard.edu/abs/2018JPhG...45e3003R}
}

@ARTICLE{Takiwaki2016,
  author = {{Takiwaki}, Tomoya and {Kotake}, Kei and {Suwa}, Yudai},
  title = "{Three-dimensional simulations of rapidly rotating core-collapse supernovae: finding a neutrino-powered explosion aided by non-axisymmetric flows}",
  journal = {Monthly Notices of the Royal Astronomical Society: Letters},
  year = {2016},
  month = {sep},
  volume = {461},
  number = {1},
  pages = {L112--L116},
  doi = {10.1093/mnrasl/slw105},
  eprint = {1602.06759},
  archivePrefix = {arXiv},
  primaryClass = {astro-ph.HE},
  adsurl = {https://ui.adsabs.harvard.edu/abs/2016MNRAS.461L.112T}
}

@ARTICLE{NagakuraAsym2019,
  author = {{Nagakura}, Hiroki and {Takahashi}, Kazuya and {Yamamoto}, Yu},
  title = "{On the importance of progenitor asymmetry to shock revival in core-collapse supernovae}",
  journal = {Monthly Notices of the Royal Astronomical Society},
  year = {2019},
  month = {feb},
  volume = {483},
  number = {1},
  pages = {208--222},
  doi = {10.1093/mnras/sty3114},
  eprint = {1811.05515},
  archivePrefix = {arXiv},
  primaryClass = {astro-ph.HE},
  adsurl = {https://ui.adsabs.harvard.edu/abs/2019MNRAS.483..208N}
}

@ARTICLE{Summa2016,
  author = {{Summa}, Alexander and {Hanke}, Florian and {Janka}, Hans-Thomas and {Melson}, Tobias and {Marek}, Andreas and {M{\"u}ller}, Bernhard},
  title = "{Progenitor-dependent Explosion Dynamics in Self-consistent, Axisymmetric Simulations of Neutrino-driven Core-collapse Supernovae}",
  journal = {The Astrophysical Journal},
  year = {2016},
  month = {jul},
  volume = {825},
  number = {1},
  eid = {6},
  pages = {6},
  doi = {10.3847/0004-637X/825/1/6},
  eprint = {1511.07871},
  archivePrefix = {arXiv},
  primaryClass = {astro-ph.SR},
  adsurl = {https://ui.adsabs.harvard.edu/abs/2016ApJ...825....6S}
}

@ARTICLE{Vartanyan2022,
  author = {{Vartanyan}, David and {Coleman}, Matthew S. B. and {Burrows}, Adam},
  title = "{The collapse and three-dimensional explosion of three-dimensional massive-star supernova progenitor models}",
  journal = {Monthly Notices of the Royal Astronomical Society},
  year = {2022},
  month = {mar},
  volume = {510},
  number = {4},
  pages = {4689--4705},
  doi = {10.1093/mnras/stab3702},
  eprint = {2109.10920},
  archivePrefix = {arXiv},
  primaryClass = {astro-ph.SR},
  adsurl = {https://ui.adsabs.harvard.edu/abs/2022MNRAS.510.4689V}
}

@ARTICLE{Roberts2016,
  author = {{Roberts}, Luke F. and {Ott}, Christian D. and {Haas}, Roland and {O'Connor}, Evan P. and {Diener}, Peter and {Schnetter}, Erik},
  title = "{General-relativistic Three-dimensional Multi-group Neutrino Radiation-hydrodynamics Simulations of Core-collapse Supernovae}",
  journal = {The Astrophysical Journal},
  year = {2016},
  month = {nov},
  volume = {831},
  number = {1},
  eid = {98},
  pages = {98},
  doi = {10.3847/0004-637X/831/1/98},
  eprint = {1604.07848},
  archivePrefix = {arXiv},
  primaryClass = {astro-ph.HE},
  adsurl = {https://ui.adsabs.harvard.edu/abs/2016ApJ...831...98R}
}

@ARTICLE{OConnorCouch2018,
  author = {{O'Connor}, Evan P. and {Couch}, Sean M.},
  title = "{Two-dimensional Core-collapse Supernova Explosions Aided by General Relativity with Multidimensional Neutrino Transport}",
  journal = {The Astrophysical Journal},
  year = {2018},
  month = {feb},
  volume = {854},
  number = {1},
  eid = {63},
  pages = {63},
  doi = {10.3847/1538-4357/aaa893},
  eprint = {1511.07443},
  archivePrefix = {arXiv},
  primaryClass = {astro-ph.HE},
  adsurl = {https://ui.adsabs.harvard.edu/abs/2018ApJ...854...63O}
}

@ARTICLE{Kuroda2018,
  author = {{Kuroda}, Takami and {Kotake}, Kei and {Takiwaki}, Tomoya and {Thielemann}, Friedrich-Karl},
  title = "{A full general relativistic neutrino radiation-hydrodynamics simulation of a collapsing very massive star and the formation of a black hole}",
  journal = {Monthly Notices of the Royal Astronomical Society: Letters},
  year = {2018},
  month = {jun},
  volume = {477},
  number = {1},
  pages = {L80--L84},
  doi = {10.1093/mnrasl/sly059},
  eprint = {1801.01293},
  archivePrefix = {arXiv},
  primaryClass = {astro-ph.HE},
  adsurl = {https://ui.adsabs.harvard.edu/abs/2018MNRAS.477L..80K}
}

@ARTICLE{Chan2018,
  author = {{Chan}, Conrad and {M{\"u}ller}, Bernhard and {Heger}, Alexander and {Pakmor}, R{\"u}diger and {Springel}, Volker},
  title = "{Black Hole Formation and Fallback during the Supernova Explosion of a 40 M$_{\odot}$ Star}",
  journal = {The Astrophysical Journal Letters},
  year = {2018},
  month = {jan},
  volume = {852},
  number = {1},
  eid = {L19},
  pages = {L19},
  doi = {10.3847/2041-8213/aaa28c},
  eprint = {1710.00838},
  archivePrefix = {arXiv},
  primaryClass = {astro-ph.SR},
  adsurl = {https://ui.adsabs.harvard.edu/abs/2018ApJ...852L..19C}
}

@ARTICLE{Nakamura2019Long,
       author = {{Nakamura}, Ko and {Takiwaki}, Tomoya and {Kotake}, Kei},
        title = "{Long-term Simulations of Multi-Dimensional Core-collapse Supernovae: Implications for Neutron Star Kicks}",
      journal = {Publications of the Astronomical Society of Japan},
         year = 2019,
        month = oct,
       volume = {71},
       number = {5},
          eid = {98},
        pages = {98},
          doi = {10.1093/pasj/psz080},
archivePrefix = {arXiv},
       eprint = {1904.08088},
 primaryClass = {astro-ph.HE},
       adsurl = {https://ui.adsabs.harvard.edu/abs/2019PASJ...71...98N}
}

@ARTICLE{Stockinger2020Long,
       author = {{Stockinger}, G. and {Janka}, H.-T. and {Kresse}, D. and {Melson}, T. and {Ertl}, T. and {Gabler}, M. and {Gessner}, A. and {Wongwathanarat}, A. and {Tolstov}, A. and {Leung}, S.-C. and {Nomoto}, K. and {Heger}, A.},
        title = "{Three-dimensional Models of Core-collapse Supernovae From Low-mass Progenitors With Implications for Crab}",
      journal = {Monthly Notices of the Royal Astronomical Society},
         year = 2020,
        month = aug,
       volume = {496},
       number = {2},
        pages = {2039--2084},
          doi = {10.1093/mnras/staa1691},
archivePrefix = {arXiv},
       eprint = {2005.02420},
 primaryClass = {astro-ph.HE},
       adsurl = {https://ui.adsabs.harvard.edu/abs/2020MNRAS.496.2039S}
}

@ARTICLE{Shibagaki2021,
  author = {{Shibagaki}, Shota and {Kuroda}, Takami and {Kotake}, Kei and {Takiwaki}, Tomoya},
  title = "{Characteristic Time Variability of Gravitational-wave and Neutrino Signals from Three-dimensional Simulations of Non-rotating and Rapidly Rotating Stellar Core Collapse}",
  journal = {Monthly Notices of the Royal Astronomical Society},
  year = {2021},
  month = {apr},
  volume = {502},
  number = {2},
  pages = {3066--3084},
  doi = {10.1093/mnras/stab228},
  eprint = {2010.03882},
  archivePrefix = {arXiv},
  primaryClass = {astro-ph.HE},
  adsurl = {https://ui.adsabs.harvard.edu/abs/2021MNRAS.502.3066S}
}

@ARTICLE{Kuroda2022QCD,
  author = {{Kuroda}, Takami and {Fischer}, Tobias and {Takiwaki}, Tomoya and {Kotake}, Kei},
  title = "{Core-collapse Supernova Simulations and the Formation of Neutron Stars, Hybrid Stars, and Black Holes}",
  journal = {The Astrophysical Journal},
  year = {2022},
  month = {jan},
  volume = {924},
  number = {1},
  eid = {38},
  pages = {38},
  doi = {10.3847/1538-4357/ac31a8},
  eprint = {2109.01508},
  archivePrefix = {arXiv},
  primaryClass = {astro-ph.HE},
  adsurl = {https://ui.adsabs.harvard.edu/abs/2022ApJ...924...38K}
}

@ARTICLE{KurodaShibata2024,
  author = {{Kuroda}, Takami and {Shibata}, Masaru},
  title = "{Numerical Relativity Simulations of Black Hole and Relativistic Jet Formation}",
  journal = {Monthly Notices of the Royal Astronomical Society: Letters},
  year = {2024},
  month = {sep},
  volume = {533},
  number = {1},
  pages = {L107--L112},
  doi = {10.1093/mnrasl/slae069},
  eprint = {2404.02792},
  archivePrefix = {arXiv},
  primaryClass = {astro-ph.HE},
  adsurl = {https://ui.adsabs.harvard.edu/abs/2024MNRAS.533L.107K}
}

@ARTICLE{Kuroda2025AIC,
  author = {{Kuroda}, Takami and {Kawaguchi}, Kyohei and {Shibata}, Masaru},
  title = "{Collapse of Rotating White Dwarfs and Multimessenger Signals}",
  journal = {Monthly Notices of the Royal Astronomical Society},
  year = {2025},
  month = {aug},
  volume = {541},
  number = {2},
  pages = {1649--1669},
  doi = {10.1093/mnras/staf1065},
  eprint = {2503.17082},
  archivePrefix = {arXiv},
  primaryClass = {astro-ph.HE},
  adsurl = {https://ui.adsabs.harvard.edu/abs/2025MNRAS.541.1649K}
}

@ARTICLE{Shibata1995,
   author = {{Shibata}, M. and {Nakamura}, T.},
    title = "{Evolution of three-dimensional gravitational waves: Harmonic slicing case}",
  journal = {\prd},
     year = 1995,
    month = nov,
   volume = 52,
    pages = {5428-5444},
      doi = {10.1103/PhysRevD.52.5428},
   adsurl = {http://ads.nao.ac.jp/abs/1995PhRvD..52.5428S}
}

@ARTICLE{Baumgarte1999,
   author = {{Baumgarte}, T.~W. and {Shapiro}, S.~L.},
    title = "{Numerical integration of Einstein's field equations}",
  journal = {\prd},
   eprint = {arXiv:gr-qc/9810065},
     year = 1999,
    month = jan,
   volume = 59,
   number = 2,
      eid = {024007},
    pages = {024007},
      doi = {10.1103/PhysRevD.59.024007},
   adsurl = {http://ads.nao.ac.jp/abs/1999PhRvD..59b4007B}
}

@ARTICLE{KurodaArcones2020,
  author = {{Kuroda}, Takami and {Arcones}, Almudena and {Takiwaki}, Tomoya and {Kotake}, Kei},
  title = "{Magnetorotational Explosion of a Massive Star Supported by Neutrino Heating in General Relativistic Three-dimensional Simulations}",
  journal = {The Astrophysical Journal},
  year = {2020},
  month = {jun},
  volume = {896},
  number = {2},
  eid = {102},
  pages = {102},
  doi = {10.3847/1538-4357/ab9308},
  eprint = {2003.02004},
  archivePrefix = {arXiv},
  primaryClass = {astro-ph.HE},
  adsurl = {https://ui.adsabs.harvard.edu/abs/2020ApJ...896..102K}
}

@article{OConnor2018,
author={O'Connor, Evan and Bollig, Robert and Burrows, Adam S. and Couch, Sean and Fischer, Tobias and Janka, Hans-Thomas and Kotake, Kei and Lentz, Eric J. and Liebend{\"o}rfer, Matthias and Messer, O. E. Bronson and Mezzacappa, Anthony and Takiwaki, Tomoya and Vartanyan, David},
title={Global comparison of core-collapse supernova simulations in spherical symmetry},
journal={Journal of Physics G: Nuclear and Particle Physics},
year={2018},
volume={45},
number={10},
pages={104001},
doi={10.1088/1361-6471aadeae}
}

@article{Bruenn1985,
author={Bruenn, Stephen W.},
title={Stellar core collapse: Numerical model and infall epoch},
journal={The Astrophysical Journal Supplement Series},
year={1985},
volume={58},
pages={771--841},
doi={10.1086/191056}
}

@article{Horowitz2002,
author={Horowitz, C. J.},
title={Weak magnetism for antineutrinos in supernovae},
journal={Physical Review D},
year={2002},
volume={65},
number={4},
pages={043001},
doi={10.1103/PhysRevD.65.043001}
}

@article{Horowitz1997,
author={Horowitz, C. J.},
title={Neutrino trapping in a supernova and the screening of weak neutral currents},
journal={Physical Review D},
year={1997},
volume={55},
number={8},
pages={4577--4581},
doi={10.1103/PhysRevD.55.4577}
}

@article{BruennMezzacappa1997,
author={Bruenn, S. W. and Mezzacappa, A.},
title={Ion screening effects and stellar collapse},
journal={Physical Review D},
year={1997},
volume={56},
number={12},
pages={7529--7547},
doi={10.1103/PhysRevD.56.7529}
}

@article{RamppJanka2002,
author={Rampp, M. and Janka, H.-T.},
title={Radiation hydrodynamics with neutrinos: Variable Eddington factor method for core-collapse supernova simulations},
journal={Astronomy \& Astrophysics},
year={2002},
volume={396},
number={1},
pages={361--392},
doi={10.1051/0004-6361:20021398}
}

@article{HannestadRaffelt1998,
author={Hannestad, Steen and Raffelt, Georg},
title={Supernova neutrino opacity from nucleon--nucleon bremsstrahlung and related processes},
journal={The Astrophysical Journal},
year={1998},
volume={507},
number={1},
pages={339--352},
doi={10.1086/306303}
}

@article{Takiwaki2014,
  author={Takiwaki, Tomoya and Kotake, Kei and Suwa, Yudai},
  title={A comparison of two- and three-dimensional neutrino-hydrodynamics simulations of core-collapse supernovae},
  journal={The Astrophysical Journal},
  year={2014},
  volume={786},
  number={2},
  pages={83},
  doi={10.1088/0004-637X/786/2/83}
}

@article{Liebendoerfer2004,
  author={Liebend{\"o}rfer, Matthias and Messer, O. E. Bronson and Mezzacappa, Anthony and Bruenn, Stephen W. and Cardall, Christian Y. and Thielemann, Friedrich-Karl},
  title={A finite difference representation of neutrino radiation hydrodynamics for spherically symmetric general relativistic supernova simulations},
  journal={The Astrophysical Journal Supplement Series},
  year={2004},
  volume={150},
  number={1},
  pages={263--316},
  doi={10.1086/380191}
}

@article{OConnor2015,
  author={O'Connor, Evan},
  title={An open-source neutrino radiation hydrodynamics code for core-collapse supernovae},
  journal={The Astrophysical Journal Supplement Series},
  year={2015},
  volume={219},
  number={2},
  pages={24},
  doi={10.1088/0067-0049/219/2/24}
}

@article{PowellEtAl2023,
  author = {Powell, Jade and M{\"u}ller, Bernhard and Aguilera-Dena, David R. and Langer, Norbert},
  title = {Three dimensional magnetorotational core-collapse supernova explosions of a 39 solar mass progenitor star},
  journal = {Monthly Notices of the Royal Astronomical Society},
  year = {2023},
  volume = {522},
  number = {4},
  pages = {6070--6086},
  doi = {10.1093/mnras/stad1292}
}

@article{BurrowsEtAl2024,
  author = {Burrows, Adam and Wang, Tianshu and Vartanyan, David},
  title = {Physical Correlations and Predictions Emerging from Modern Core-collapse Supernova Theory},
  journal = {The Astrophysical Journal Letters},
  year = {2024},
  volume = {964},
  number = {1},
  pages = {L16},
  doi = {10.3847/2041-8213/ad319e}
}

@article{VarmaMueller2026,
  author = {Varma, Vishnu and M{\"u}ller, Bernhard},
  title = {Resolution dependence in magnetohydrodynamic simulations of neutrino-driven core-collapse supernovae},
  journal = {Monthly Notices of the Royal Astronomical Society},
  year = {2026},
  volume = {548},
  number = {2},
  pages = {stag626},
  doi = {10.1093/mnras/stag626}
}

@article{Shankar2026,
 author={Shankar, S. and M{\"o}sta, P. and Haas, R. and Schnetter, E.},
 title={3D full-{GR} simulations of magnetorotational core-collapse supernovae on {GPU}s: a systematic study of rotation rates and magnetic fields},
 journal={Monthly Notices of the Royal Astronomical Society}, year={2026}, volume={548}, number={2}, pages={stag646},
 doi={10.1093/mnras/stag646}}

@article{Palenzuela2025,
 author={Palenzuela, C. and Bezares, M. and Liebling, S. and Schianchi, F. and Abalos, J. F. and Aguilera-Miret, R. and Bona, C. and Carretero, J. A. and Massò, J. and Smith, M. P. and Amponsah, K. and Kornet, K. and Mi{\~n}ano, B. and Pareek, S. and Radia, M.},
 title={{MHDuet}: a high-order general relativistic radiation {MHD} code for {CPU} and {GPU} architectures},
 journal={Classical and Quantum Gravity}, year={2025}, volume={42}, number={24}, pages={245005},
 doi={10.1088/1361-6382/ae255e}}

@article{Endeve2026,
 author={Endeve, E. and Mewes, V. and Harris, J. A. and Laiu, M. P. and Chu, R. and Fromm, S. A. and Mezzacappa, A. and Messer, O. E. B. and Hix, W. R. and Bruenn, S. W. and Lentz, E. J. and Weide, K. and Cardall, C. Y. and Almgren, A. S. and Dubey, A. and Couch, S. M. and M{\"o}sta, P. and Willcox, D. E.},
 title={{thornado+Flash-X}: a hybrid discontinuous Galerkin--implicit-explicit and finite-volume framework for neutrino-radiation hydrodynamics in core-collapse supernovae},
 journal={The Astrophysical Journal Supplement Series}, year={2026}, volume={284}, number={2}, pages={40},
 doi={10.3847/1538-4365/ae57ad}}
\label{lastpage}
\end{document}